\documentclass[reprint,pra,longbibliography,superscriptaddress,nofootinbib,aps]{revtex4-2}
\usepackage{lipsum}
\usepackage{fancyhdr}
\usepackage{graphicx}
\usepackage[caption=false]{subfig}
\usepackage{braket}
\usepackage[dvipsnames]{xcolor}
\usepackage{color,soul}
\usepackage{mathtools}
\usepackage{amsmath}
\usepackage{amssymb}
\usepackage{hyperref}
\usepackage{esvect}
\usepackage{float}
\usepackage[normalem]{ulem}

\graphicspath{{./PaperFigures/}}

\allowdisplaybreaks

\newcommand{\edit}[1]{{\color{black} #1}}

\hypersetup{
    colorlinks=true,
    linkcolor=blue,
    filecolor=magenta,      
    urlcolor=cyan,
    citecolor={blue},
    }

\begin{document}

\title{Resonance fluorescence of an artificial atom with a time-delayed coherent feedback}

%Is there a reason these two affiliations are up here? It puts them out of order below
% MUST BE IN ORDER OF AUTHORS - SO COMMENTIN GOUT - THAT IS A LOT OF ADDRESSES!!
% \affiliation{Department of Physics, National Tsing Hua University, Hsinchu 30013, Taiwan}
% \affiliation{Department of Physics, City University of Hong Kong, Kowloon, Hong Kong SAR, China}

\author{C.-Y. Chen}
\thanks{Authors C.-Y. Chen and G. Crowder contributed equally. \\ \\ Corresponding authors:}
\affiliation{Department of Physics, National Tsing Hua University, Hsinchu 30013, Taiwan}
\author{G. Crowder}
\thanks{Authors C.-Y. Chen and G. Crowder contributed equally. \\ \\ Corresponding authors:}
\affiliation{Department of Physics, Queen's University, Kingston, Ontario, Canada, K7L 3N6}
\affiliation{Department of Physics and Nexus for Quantum Technologies Institute, University of Ottawa, Ottawa, Ontario, Canada, K1N 6N5}
\author{Z.-Q. Niu}
\affiliation{State Key Laboratory of Materials for Integrated Circuits, Shanghai Institute of Microsystem and Information Technology (SIMIT),Chinese Academy of Sciences, Shanghai 200050, China}
\affiliation{ShanghaiTech University, Shanghai 201210, China}
\author{P. Y. Wen}
\affiliation{Department of Physics, National Chung Cheng University, Chiayi 621301, Taiwan}
\author{Y.-H. Lin}
\affiliation{Department of Physics, National Tsing Hua University, Hsinchu 30013, Taiwan}
\affiliation{Center for Quantum Technology, National Tsing Hua University, Hsinchu 30013, Taiwan}
\author{J.-C. Chen}
\affiliation{Department of Physics, National Tsing Hua University, Hsinchu 30013, Taiwan}
\affiliation{Center for Quantum Technology, National Tsing Hua University, Hsinchu 30013, Taiwan}
\author{Z.-R.~Lin}
%\email{zrlin@mail.sim.ac.cn}
\affiliation{State Key Laboratory of Materials for Integrated Circuits, Shanghai Institute of Microsystem and Information Technology (SIMIT),Chinese Academy of Sciences, Shanghai 200050, China}
\author{F. Nori}
\affiliation{Center for Quantum Computing, RIKEN, Wakoshi, Japan}
\affiliation{Physics Department, University of Michigan, Ann Arbor, USA}
\author{S. Hughes}
\email{shughes@queensu.ca}
\affiliation{Department of Physics, Queen's University, Kingston, Ontario, Canada, K7L 3N6}
\author{I.-C. Hoi}
\email{iochoi@cityu.edu.hk}
\affiliation{Department of Physics, City University of Hong Kong, Kowloon, Hong Kong SAR, China}
\affiliation{Department of Physics, National Tsing Hua University, Hsinchu 30013, Taiwan}

%\footnote{test}
\date{\today}

\begin{abstract}
The model of light-matter interaction in quantum electrodynamics typically relies on the Markovian approximation, which assumes that the system's future evolution depends solely on its current state, effectively treating it as a ``memoryless" process. However, this approximation is not valid in scenarios when retardation effects are significant. These memory and retardation effects have the potential to improve existing quantum technologies (e.g., large-scale quatnum networks, quantum information processing) and unlock new phenomena for future applications. In this work, we show theory and experiments of a time-delayed coherent feedback system using a transmon artificial atom (treated as a qubit) embedded in a superconducting circuit waveguide, \edit{in both linear and nonlinear excitation regimes}. By using a feedback loop with a delay time comparable to the qubit relaxation time, pronounced non-Markovian effects appear in the dynamics of the qubit evolution. We also show how the resonance fluorescence spectrum, including elastic and inelastic scattering (such as the well-known Mollow triplet), can be significantly modified through the interaction between the qubit and feedback loop to show genuine non-Markovian \edit{and quantum nonlinear} phenomena that cannot be explained with instantaneous coupling parameters. 
\edit{This work presents the first experimental report of Mollow triplets in the non-Markovian regime.}

\end{abstract}

\maketitle

The interaction between a two level system (TLS) and an electromagnetic field is a foundation in quantum optics, with applications in fundamental optical physics and quantum technologies~\cite{kimble2008quantum}. 
This system has led to a diverse range of intriguing optical physics, such as photon-antibunching~\cite{kimble1977photon}, single photon sources~\cite{eisaman2011invited}, squeezed light~\cite{toyli2016resonance} and Mollow triplets~\cite{lu2021characterizing,astafiev2010resonance,kim2014resonant}. Under resonant (or near-resonant) driving, the TLS exhibits resonance fluorescence~\cite{Kimble1976}, where the emitted spectrum~\cite{lakowicz2006principles} can result in the distinctive Mollow triplet with strong driving~\cite{Mollow1969}, featuring a central peak at the frequency of the TLS and two side peaks offset by the Rabi frequency of the pump field. Fluorescence spectra have been demonstrated in various physical systems, including natural atoms~\cite{masters2023simultaneous,schuda1974observation}, chiral artificial atoms~\cite{joshi2023resonance,redchenko2023tunable}, quantum dots~\cite{nick2009spin,Ulhaq2013,kim2014resonant}, and superconducting circuits~\cite{astafiev2010resonance,gasparinetti2019two,lu2021characterizing}. These platforms have also been studied with exotic driving fields such as a squeezed vacuum~\cite{toyli2016resonance}, a dynamical drive~\cite{ruan2024dynamics,boos2024signatures,liu2024dynamic} and entangled photons~\cite{lopez2024entanglement}.

Superconducting circuits provide an exciting platform for studying non-Markovian time delay effects on resonance fluorescence in a single artificial atom waveguide system~\cite{blais2021circuit,kockum2019quantum,gu2017microwave}. A major advantage of these circuits is their low loss and the sub-wavelength control of the light-matter interaction made possible by tuning the qubit's frequency within the circuit. In recent years, considerable progress has been made toward achieving strong coupling between superconducting artificial atoms and propagating microwave photons~\cite{kockum2018decoherence,kannan2020waveguide,liul2023coherent,kurpiers2018deterministic,joshi2023resonance}. Thus superconducting artificial atoms and circuits make excellent candidates for exploring waveguide quantum electrodynamics (QED), with precise tunability.

Using superconducting QED systems, a wide range of quantum field phenomena have been demonstrated. These include resonance fluorescence~\cite{astafiev2010resonance}, photon routing~\cite{hoi2011demonstration}, non-classical microwaves~\cite{hoi2012generation}, the cross-Kerr effect between two fields~\cite{hoi2013giant}, lifetime control of artificial atoms~\cite{hoi2015probing}, amplification without population inversion~\cite{wen2018reflective}, the collective Lamb shift between two superconducting qubits~\cite{wen2019large,lin2019scalable}, giant atoms~\cite{kannan2020waveguide}, photon-mediated interactions between atoms~\cite{van2013photon,kannan2023demand}, and deterministic photon loading~\cite{lin2022deterministic,cheng2024tuning}. These 
%groundbreaking 
experiments have inspired new directions for exploiting finite time delay effects of a single atom within the framework of waveguide-QED.
%using superconducting circuits.

In the realm of waveguide-QED, theoretical work has predicted that a time-delayed coherent feedback can significantly alter the characteristics of resonance fluorescence by introducing new resonances and changing the photon output statistics~\cite{Dorner2002,Pichler2016,zhang2017quantum,Lu2017,Crowder2022}. In contrast to measurement-based feedback~\cite{Munoz-Arias2020,Grigoletto2021,Borah2021}, {\it coherent feedback} is included at the system level to avoid measurements (and the resulting wavefunction collapse), and thus one can preserve the coherence of the system dynamics, a key resource in quantum technology applications. Furthermore, a finite time delay is also included with the feedback, so that the system output returns after a non-trivial delay time, which introduces non-Markovian (i.e., non-instantaneous, retardation) dynamics to the system evolution~\cite{Zhang2012,Tufarelli2013,Carmele2013,Grimsmo2015,Nemet2016,Chen2016,Pichler2017,Whalen2017,andersson2019non,Calajo2019,Crowder2020,Lu2020,Regidor2021a,Crowder2022,Barkemeyer2022,Crowder2024,Vodenkova2024,ferreira2021collapse,Ferreira2024,bienfait2019phonon,zhong2019violating,odeh2025non}. This is a challenging dynamic to simulate because conventional model approaches, such as the master equation, use a Markov approximation which fails in the presence of a time-delayed coherent feedback~\cite{QuantumNoise,Stenius1996}. Here, we use a quantum trajectory discretized waveguide (QTDW) method which includes the feedback at the system level to be able to describe the non-Markovian system with Markovian equations~\cite{Whalen2019,Regidor2021a,Crowder2022,Crowder2024}.

In this work, we use a superconducting circuit and transmon artificial atom platform (treated as a two level qubit) to implement a coherent feedback loop with a delay time on the order of the qubit's radiative lifetime. By controlling the qubit and pump frequencies, the reflection coefficient is recorded to characterize the qubit and feedback loop. \edit{While retardation effects have been previously studied in the linear regime~\cite{ferreira2021collapse,bienfait2019phonon,zhong2019violating,odeh2025non}, here we study the quantum nonlinear behavior of this artificial atom by measuring and simulating the resonance fluorescence spectrum under a strong resonant pump and Mollow triplets} as a function of pump power at two contrasting round trip phase changes, which \edit{qualitatively} agree. These results highlight a new quantum optics regime where 
a time-delayed feedback changes the well-known Mollow triplet, and we show simulation results which highlight that this regime shows a pronounced non-Markovian response.

These feedback effects go to the root of time-delayed non-instantaneous coupling which can produce new and exotic quantum optical states. For example,  the time delay can be strategically chosen for the returning field to interfere (constructively or destructively) with the qubit emission during an excitation protocol, with applications to improving performance of single photon sources~\cite{Crowder2024} or generating new resonances~\cite{Crowder2022}. Also, the time delay can be lengthened to be longer than an individual pulse to interfere with the qubit output between sequential excitations such as in cluster state generation protocols~\cite{Ferreira2024}. Here, the time delay is on the order of the qubit lifetime and sets up nodes where the Mollow triplet sidepeaks are suppressed and can introduce new resonances through \textit{dressing} of the Fabry-P\'{e}rot modes beyond the three peaks of the Mollow triplet.

\section*{Time-Delayed Coherent Feedback System}

\subsection*{Theoretical Model}

\begin{figure}[t]
    \centering
    \includegraphics[width=1\columnwidth]{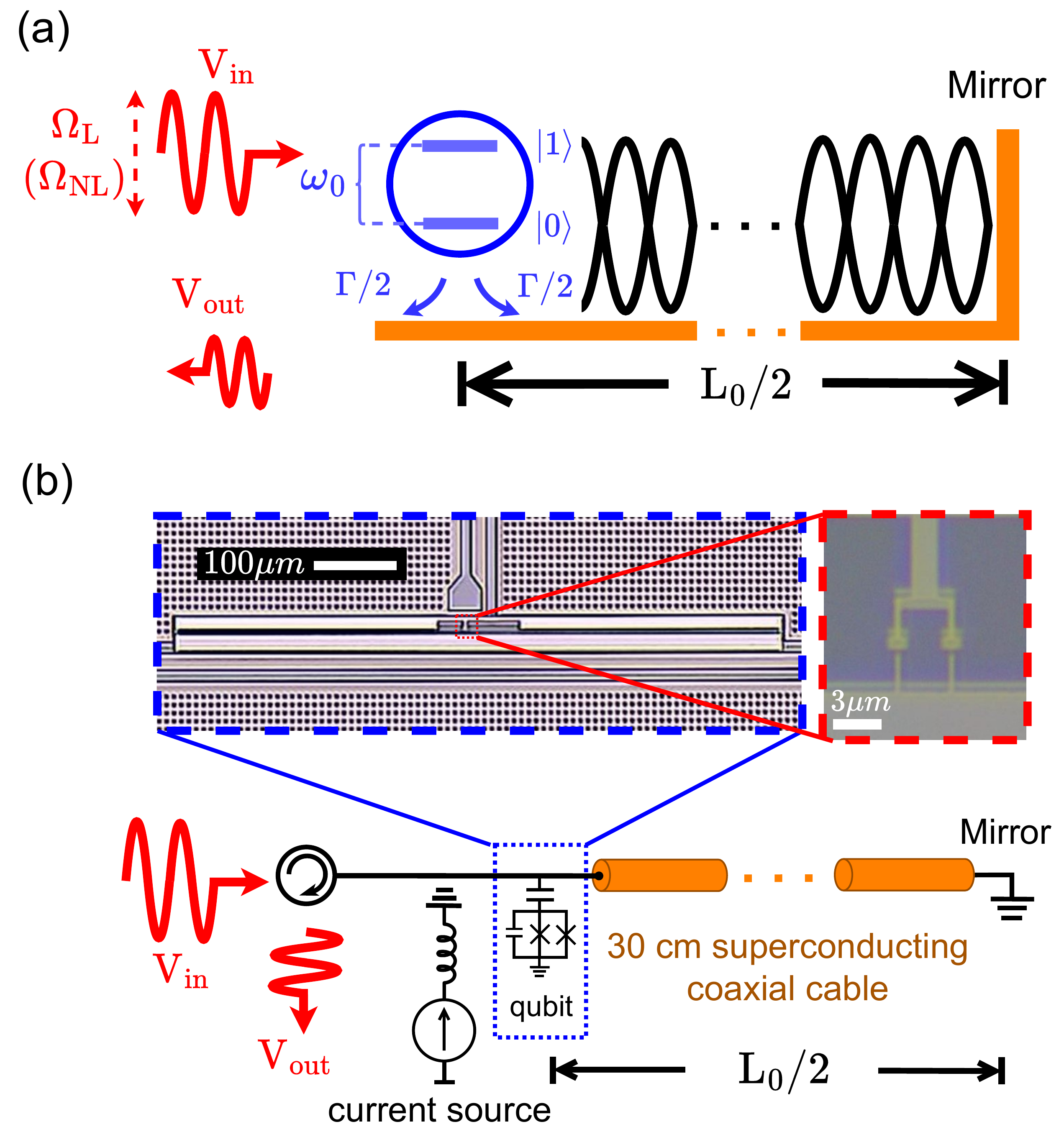}
    \caption{(a) Schematic of the qubit-mirror in a quasi-1D system. The round trip length is $L_{0}$ and the qubit couples to the waveguide symmetrically with total relaxation rate $\Gamma$. The qubit is pumped with frequency $\omega_{\rm{p}}$ and Rabi frequency \edit{ $\Omega_{\rm{L}}(\Omega_{\rm{NL}})$} directly through the waveguide. (b) Optical microscope images of the transmon qubit~\cite{koch2007charge} and the experimental setup. The blue dashed rectangle shows the qubit capacitively coupled to a 1D transmission line and the red dashed square shows the superconducting quantum interference device (SQUID) with a tunable frequency through the applied external magnetic flux \edit{with the superconducting magnet. The structure above the red dashed square is the flux line but unused in this experiment.} One end of the transmission line couples to the feedback loop terminated by a connection to the grounding plane that acts as the mirror, and the material of the superconducting coaxial cable is NbTi. The other end connects to a three port circulator through which the input signal enters the waveguide and the output signal is detected. The details of the \edit{waveguide-mirror system and the} setup can be found in \edit{Fig.~S3 and Fig.~S2 of Sec. II of the SM~\cite{SuppMaterial}, respectively}.
    %To realize the setup, we have the Transmon qubit capacitively coupled to the 1D TL, shown in the green dash square optical microscopic image, and in the red dash square, the SQUID shows the frequency could be tuned by applying external flux $\Phi$. The characteristic impedance of the TL is $Z_0\approx50 \Omega$. One of the TL port couple to the 0.3 m superconducting coaxial cable, also with characteristic impedance $Z_0\approx50\Omega$ at the low temperature. At the end of the cable is grounding by the shorted terminator. The other port of the TL, we connect it to the three port circulator. In here, we input the signal into the waveguide, to probe and pump the qubit, and also detect the reflected signal.
    }
    \label{Schematic}
\end{figure}

As illustrated in Fig.~\ref{Schematic}(a), a transmon qubit couples bidirectionally to the waveguide with a total relaxation rate of $\Gamma$ in an open transmission line and a resonance frequency of $\omega_0$. The presence of a mirror at the end of the waveguide allows the propagating fields (microwaves) to interact with the qubit after each round trip with a delay of $\tau = L_0/c(\omega_0)$, where $L_0$ is the round trip length of the feedback loop and $c(\omega_0)$ is the group velocity of the waveguide mode with reference frequency $\omega_0$~\cite{hoi2015probing}. The feedback loop also introduces a round trip phase change $\phi = \phi_{\rm{M}} + \omega_0 \tau$, where $\phi_{\rm{M}}$ is the mirror phase change, in this case $\phi_{\rm{M}} = \pi$ and we calculate $\phi$ relative to the antinode when $\omega_0/2\pi = 4.9$ GHz. 

The qubit is excited by a continuous-wave (CW) field through the waveguide with frequency $\omega_{\rm{p}}$. Since the pump is introduced through the waveguide, it will set up a standing wave due to its reflection off the mirror, leading to an effective Rabi frequency of the form
\begin{equation}
\Omega_{\rm{eff}} = 
\Omega_{\rm{NL}} (1 + e^{i \phi_{\rm{p}}}),
\end{equation}
with $\Omega_{\rm{NL}}$ the Rabi frequency in the non-linear regime and $\phi_{\rm{p}} = \phi_{\rm{M}} + \omega_{\rm{p}} \tau$.
In particular, when $\phi_{\rm{p}} = 0 \,\,(\pi)$, the qubit is located at an antinode (node) of the pump's electromagnetic field. We also include pure dephasing effects in the qubit with rate $\gamma'$, an important dissipation channel to include in order to compare simulated observables with real experiments, and a phenomenological radiative loss term with rate $\gamma_{\rm{L}}$ dependent on $\phi$ (for example with $\gamma_{\rm{L}} \approx 0$ when $\phi = 0$ and $\gamma_{\rm{L}} = 0.075\Gamma$ when $\phi = -0.632\pi$).

We highlight that if the feedback loop does not introduce a long enough delay, the dynamics will remain approximately Markovian (mathematically $\tau \rightarrow 0$), and the loop will only act to introduce a phase change $\phi_{\rm{M}}$. However, when $\tau$ is on the order of the qubit lifetime, non-Markovian effects will be introduced and this typically occurs when $\edit{\Gamma}\tau > 0.1$.

\edit{Two main extensions have been made here to solve coherent state driving in the presence of the mirror, in both the linear and nonlinear regimes. In the linear (weak field) regime, the input bins in the QTDW model representing the incoming waveguide field enter as a superposition between the vacuum state and the single photon state:
\begin{equation}
    \ket{\psi'} = \sqrt{1 - \Omega_{\rm{L}} dt} \ket{0}_{N-1} + \sqrt{\Omega_{\rm{L}} dt} \ket{1}_{N-1},
\end{equation}
which introduces a photon flux through the waveguide of $\Omega_{\rm{L}}dt$ with $\Omega_{\rm{L}}$ with Rabi frequency in the linear regime. This is valid when the total flux present in the loop is less than the truncation of the number of photons in the feedback loop, $\Omega_{\rm{L}} \tau < 2$, which is the case when dealing with results in the relatively weak pumping regime. The reflection coefficient in the presence of this coherent field is
\begin{equation}
    r = \frac{e^{i \delta \tau}\braket{B_0}_{\rm{ss}}}{\Omega_{\rm{L}} dt},
\end{equation}
where $\delta = \omega_0 - \omega_{\rm{p}}$ is the detuning between qubit and pump and  $\braket{B_0}_{\rm{ss}}$ is the expectation of the final waveguide bin (introduced in the QTDW model) in the steady state.

In the nonlinear regime, the assumption of a weak pump fails and instead the coherent portion and incoherent fluctuations of the waveguide field are treated separately. To solve this problem, we derive an effective coherent pump term in the presence of feedback that is included in the Hamiltonian, and then treat the incoherent fluctuations with the QTDW model. The incoherent spectrum is given by 
\begin{equation}
    \begin{aligned}
        S_{\rm incoh}^{\rm out}(\omega) ={} & \int_0^{\infty} d t_2 \rm{Re} \left[ e^{i(\omega-\omega_{\rm p}) t_2} \times \right.\\
        & \hspace{-1cm} \left. \left( \braket{B_0^{\dagger}(t_2) B_0 (0)}_{\rm ss} -\braket{B_0^{\dagger}(0)}_{\rm ss} \braket{B_0(0)}_{\rm ss} \right) \right],
    \end{aligned}
\end{equation}
where $\braket{B_0^{\dagger}(t_2) B_0 (0)}_{\rm ss}$ is the unnormalized first-order auto-correlation function which we calculate following the technique described in~\cite{Crowder2022}.

For additional details, including the  Hamiltonian terms in both regimes, and an explanation of the QTDW model used to simulate the qubit and feedback system, see Sec. I of the Supplemental Material (SM)~\cite{SuppMaterial}.}

\subsection*{Experimental Methods}

Figure~\ref{Schematic}(b) shows an optical image of our device, a superconducting transmon qubit~\cite{koch2007charge} strongly coupled to a one dimensional (1D) open transmission line~\cite{astafiev2010resonance,hoi2011demonstration} with characteristic impedance $Z_{0} \approx 50 \,\, \Omega$. This system acts as a near perfect waveguide creating an excellent platform to study waveguide-QED experiments. A unique property of these qubits is the ability to control the transition energy determined by
\begin{equation}
    \hbar \omega_{0} (\Phi_{\rm{ext}}) \approx \sqrt{8 E_{J}(\Phi_{\rm{ext}}) E_{C}} - E_{C},
    \label{qubit}
\end{equation}
where $E_C$ is the charging energy, a constant given by $E_{C} = e^2/2C_{\Sigma}$, where $C_{\Sigma}$ is the total capacitance of the qubit. The Josephson energy, $E_J(\Phi_{\rm{ext}})$, is determined by $E_{J}(\Phi_{\rm{ext}}) = E_{J,0}|\cos(\pi \Phi_{\rm{ext}} / \Phi_{0})|$ where $\Phi_{\rm{ext}}$ is the applied external magnetic flux, which can be controlled by the current, $I$, and $\Phi_{0} = h / (2e)$ is the flux quantum. %The josephson energy $E_{J}(\Phi) = E_{J}|\cos(\pi\Phi/\Phi_{0})|$ can be controlled by applying the external flux $\Phi$ and $\Phi_0 = h/2e$ is the megnetic flux quantum.
In this experiment, the characterization of the qubit is shown in Fig.~\edit{S4 of the SM~\cite{SuppMaterial}} and the extracted parameters are given in Table~\edit{S1 of the SM~\cite{SuppMaterial} } (the extraction technique can be found in Refs.~\cite{wen2018reflective,aziz2025nearly}), which introduces an anharmonicity allowing the qubit to be treated as a TLS with ground state $\ket{0}$ and first excited state $\ket{1}$.

In order to access the non-Markovian regime with time-delayed feedback, we need a time delay between the qubit and mirror on the order of the qubit lifetime ($1/\Gamma$). To achieve this, the transmission line is connected to a superconducting coaxial cable which is truncated by connecting the center conductor to the grounding plane with a shorted terminator, contributing the mirror phase change: $\phi_{\rm{M}}=\pi$. This superconducting cable has $L_0/2 \approx 0.3 \,\, \rm{m}$, corresponding to a round trip delay time of $\tau= 3.63 \,\, \rm{ns}$. The characterization method is \edit{discussed in Sec. IV of the SM ~\cite{SuppMaterial}}. Although the qubit-mirror distance is fixed, we can change the round trip phase change, $\phi$, by controlling $\omega_0$ and corresponding effective wavelength. This allows the qubit to be tuned to the node or antinode of the waveguide field, i.e., a round trip phase change that varies from $\pi$ to 0. Further details of the experimental setup are available in \edit{Sec. II to IV of the SM~\cite{SuppMaterial}}.
%Under this setup, the total output field can only propagate in one direction.

%\begin{equation}
%\begin{split}
%{\phi}={2\times[2\pi L_{0}/2\lambda] + \pi}
%\end{split}
%\end{equation}

\section*{Elastic Scattering of an Incident Microwave Field - Linear Regime}

\begin{figure}
    
    \includegraphics[width=1\columnwidth]{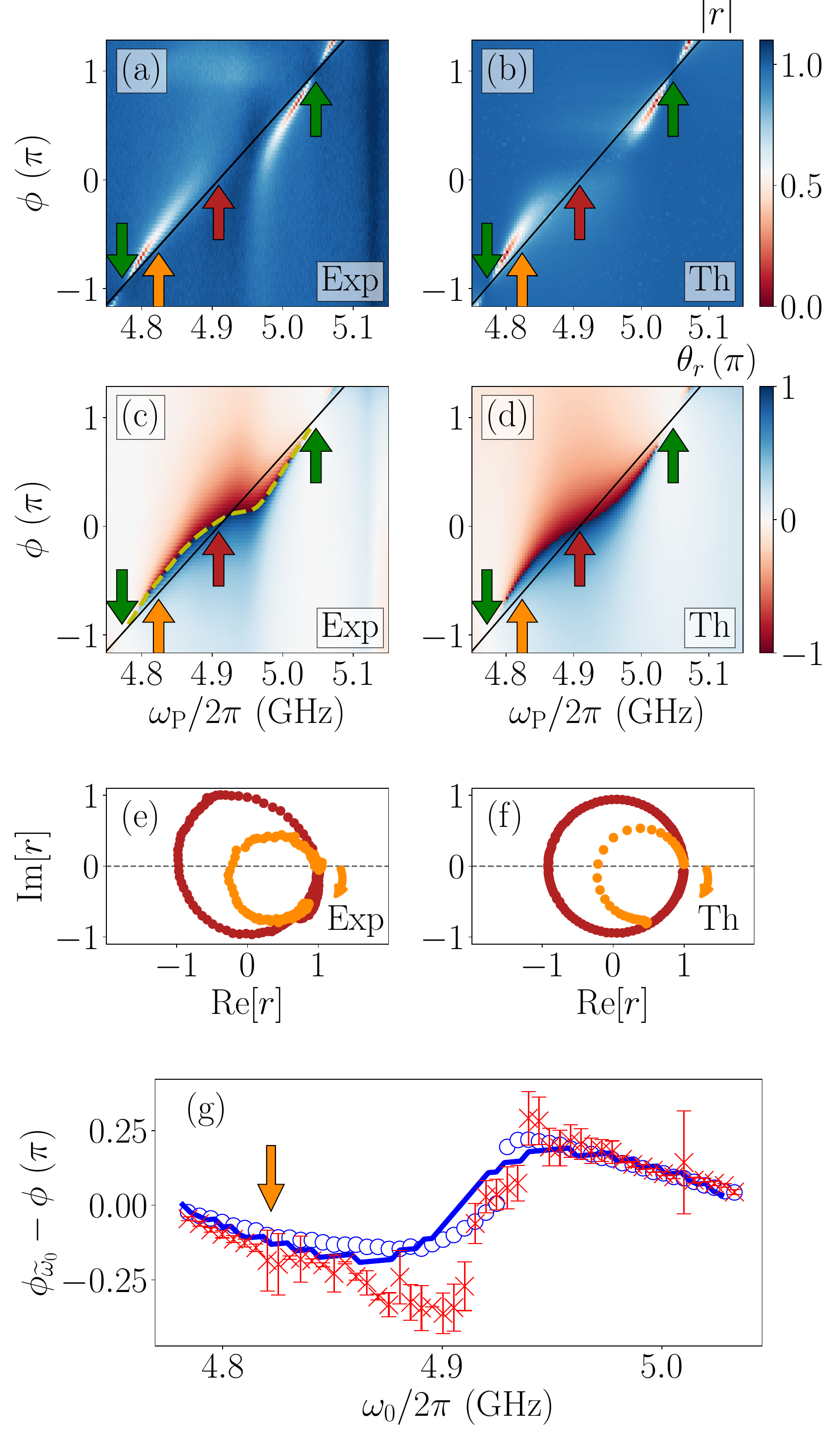}
    
    \caption{Single-tone spectroscopic measurement of the reflection coefficient $r$ as a function of $\phi$ and $\omega_{\rm{p}}$ using a weak pump ($\edit{\Omega_{\rm{L}}} \ll \gamma_{\rm{d}}$). (a) and (b) show the magnitude response, $|r|$, and (c) and (d) show the phase response, $\theta_r$, of the experimental and simulated $r$, respectively (the full spectroscopy is shown in Fig.~\edit{S4 of the SM~\cite{SuppMaterial}}. The node and antinode frequencies of the pump field are indicated by the green and red arrows respectively. The solid black line in these graphs indicate the expected qubit bare resonance frequency $\omega_0$ without coupling to the continuous photonic mode, calculated by the extracted anharmonicity and maximum qubit frequency. \edit{The yellow dashed curve shows the frequency shift in experiment.} (e) and (f) respectively show experimental and simulated line cuts from (a)-(d) of $r$ in the complex plane for fixed $\omega_0$ indicated by the red ($\omega_0/2\pi = 4.9 \,\, \rm{GHz}$, $\phi = 0$) and orange ($\omega_0/2\pi = 4.822 \,\, \rm{GHz}$, $\phi = -0.632\pi$) arrows. The orange curved arrows show the tilted angle of the complex plane. (g) shows $(\widetilde\omega_0 - \omega_0)\tau$ of the experimental (blue data points) and simulated (blue solid curve) results as well as the tilted angle $\phi_{\widetilde\omega_{\rm{0}}}-\phi$ directly extracted from the complex plane (red crosses data points) versus $\omega_0$. The orange arrow points out the corresponding $\phi_{\widetilde\omega_{\rm{0}}}-\phi$ of the orange complex plane in (e) and (f). The disagreement between data and theory around $4.9\,\, \rm{GHz}$ is likely due to the qubit coupling to a stray resonance~\cite{en1}. 
    %Figures (a) and (b) display the magnitude and phase response, respectively. The white dashed line indicates a node distance of 270 MHz, generated by the mirror at the end of the transmission line (TL). When the qubit frequency corresponds to $L=\Lambda n/2$, the qubit resides at the node of the field and is therefore hidden from the probe, resulting in no observed signal. Figure (c) shows the complex plane indicated by the red (orange) arrows in (a) and (b). The solid lines represent the circle fits from which we extract the decay rate $\gamma$, decoherence rate, and resonance frequency. We derive the $T_1$ value from the inverse lifetime $T_1 = 2/\Gamma$. The resonance frequency is approximately 4.9 (4.806) GHz, with a relaxation rate of 96 (7.2) MHz, and T1 is 1.83 (20) ns. The fitting is performed without considering the non
    }
    \label{Flux_dp}
\end{figure}

Before studying the nonlinear regime of the resonance fluorescence, we investigate the elastic scattering of the system with a weak pump ($ \edit{\Omega_{\rm{L}}} \ll \gamma_{\rm{d}}$, where $\gamma_{\rm{d}} \edit{= \gamma' + \widetilde{\Gamma} /2}$ is the \edit{measured} decoherence rate \edit{and $\widetilde{\Gamma}$ is the measured relaxation rate}, which approximately reflects the loss of quantum coherence from both energy relaxation and pure dephasing), to characterize the qubit response in the linear regime. This is done using a Vector Network Analyzer (VNA) to send the input voltage, $V_{\rm{in}}$, with a frequency around $\omega_0$, sent in through the waveguide and the outgoing voltage, $V_{\rm{out}}$, measured to retrieve the reflection coefficient $r=\langle V_{\rm{out}}\rangle /\langle V_{\rm{in}}\rangle$. By controlling the external magnetic flux, $\Phi_{\rm{ext}}$, via applying current, $I$, the qubit frequency is swept from $4.75\,\, \rm{GHz} < \omega_0/2\pi < 5.08\,\, \rm{GHz}$ (equivalent to a range of $-1.16\pi <\phi < 1.24\pi$) and the pump frequency is swept from $4.75\,\, \rm{GHz} < \omega_{\rm{p}}/2\pi < 5.15\,\, \rm{GHz}$. This maps out a parameter space containing the 18th and 19th nodes of the standing waveguide mode.

Figures~\ref{Flux_dp}(a)–(d) show the measured and simulated reflection coefficients as functions of the pump frequency, $\omega_{\rm{p}}$, and the round trip phase change, $\phi$. The offset from the resonance condition between the pump and qubit (black line) is caused by a feedback induced shift from $\omega_0$ to $\widetilde\omega_0$~\cite{koshino2012control,wu2024microwave}. The qubit will be aligned with a node in the standing waveguide mode when $\omega_0 / 2\pi = 4.772$ GHz or 5.047 GHz corresponding to $\phi = -\pi$ or $\pi$, respectively. 

%The parameter sweep across $\omega_0$ and $\omega_{\rm{p}}$ then covers over a full period in both $\phi$ and $\phi_{\rm{pump}}$ to see the complete range of the system response. Additionally, Figs.~\ref{Flux_dp}(e) and (f) show the reflection coefficient in the complex plane as the probe field frequency changes for $\omega_0/2\pi = 4.9\,\, \rm{GHz}$ and $\omega_0/2\pi = 4.806\,\, \rm{GHz}$ respectively, corresponding to $\phi = 0$ and $\phi = -3\pi/4$. 
%The radiative shift between the signal of $\omega_{0}$ and the black line due to the qubit couples to the continuous photonic mode ~\cite{koshino2012control}. We observed the full period of the $\phi$ from $\pi$ to $-\pi$, corresponding to the node frequency we observed at $4.772$ and $5.047\, \, \rm{GHz}$. We plotted the reflection coefficients $\rm{r}$ in a complex-plane at two set frequencies, $\omega_0/2\pi = 4.9\,\, \rm{GHz}$ and $\omega_0/2\pi = 4.806\,\, \rm{GHz}$ corresponding to $\phi = 0$ and $\phi = -3\pi/4$ respectively, in Figs.~\ref{Flux_dp}(e) and (f). 

At $\phi = 0$, when the qubit is at the antinode of the standing waveguide mode, indicated by the red arrow in Figs.~\ref{Flux_dp}(a)-(d), the interaction between the qubit and feedback results in enhanced spontaneous emission. At this point, the reflection spectroscopy is dominated by coherent scattering (elastic scattering), so the qubit fully reflects the resonant pump field without sending energy into the feedback loop, giving $|r| \approx 1$, but the phase is flipped indicating the scattering from the qubit. Perfect reflection also occurs when $\phi = \pm \pi$, but without a phase change. This is because the qubit is now at a node of the standing wave and is invisible to the incoming pump field. Thus, no scattering from the qubit occurs, indicating that the qubit is completely transparent to the waveguide field and perfect reflection from the mirror occurs.

Between $\phi = 0$ and $\phi = \pm \pi$ when the probing frequency is on resonance, incoherent scattering (i.e., from phase jumps due to pure dephasing) by the qubit produces a dip in the reflected field when $\omega_{\rm{p}} \approx \widetilde\omega_0$. As $\phi$ changes from $0$ to $\pm\pi$, the feedback modified effective relaxation rate, $\widetilde\Gamma$, will change from its maximum to 0 because the qubit is effectively moving from an antinode to a node of the standing wave of the waveguide mode. When $\widetilde\Gamma$ is close to $\gamma_{\rm L}/2 + \gamma'$, part of the incident photons will be scattered incoherently through pure dephasing or intrinsic loss, which means $|r|$ can drop to 0~\cite{cheng2024group}. Then finally, the dip becomes narrower and eventually disappears, with no phase change, because the qubit sits at a node and is not pumped by the standing wave mode when $\phi_{\rm{p}}$ approaches $\pm\pi$.

%Between $\phi = 0$ and $\phi = \pm \pi$ when the probe is on resonance, there is still coherent scattering of the probe field but it is no longer perfect as the qubit does not see the total probe field and is only driven by at $\Omega_{\rm eff}$. This in turn causes interference between the scattered field and the returning probe field that passes around the feedback loop. When this interference is destructive, such as between $\phi = \pm \pi$ and $\phi = \pm \pi /2$ the reflection coefficient drops and the perfect reflection is lost.

Although our system is in the non-Markovian regime, we can still use the experimental response in Figs.~\ref{Flux_dp}(e) to extract information for $\widetilde\Gamma$ and $\gamma_{\rm{d}}$ by the circle fit technique~\cite{probst2015efficient}. \edit{The extracted values serve as a baseline for our non-Markovian QTDW simulation.} The corresponding fitting function is 
\begin{equation}
    r(\omega_{\rm{p}}) = 1 - \frac{i\widetilde\Gamma(\omega_{\rm{p}})}{\Delta + i\gamma_{\rm{d}}(\omega_{\rm{p}})},
    \label{ref_coeff_eqn}
\end{equation}
where $\Delta\,\, =\,\,\omega_{\rm{p}} - \widetilde\omega_0$. Figures~\ref{Flux_dp}(e) and (f) show the reflection coefficient in the complex plane as the pump field frequency changes for $\widetilde\omega_{\phi=0}/2\pi = 4.9\,\, \rm{GHz}$ and $\widetilde\omega_{\phi=-0.632\pi}/2\pi = 4.806\,\, \rm{GHz}$, respectively corresponding to $\phi = 0$ (red circle) and $\phi = -0.632\pi$ (orange circle). We compared the complex plane of the reflection coefficient with the experimental data, and demonstrate an excellent fit.
%simulation result, and the circle fit result.

For $\phi = 0$, in Fig.~\ref{Flux_dp}(e), the extracted parameters from the qubit's reflection response are in Table \edit{S2 of the SM~\cite{SuppMaterial}}.
%we estimate a decoherence rate of $\gamma_{\rm{d}} / 2\pi = 50.01 \,\, {\rm{MHz}}$ and the effective relaxation rate is $\widetilde\gamma_{\phi = 0} / 2\pi = 96.2 \,\, {\rm{MHz}}$. 
Since the qubit is at an antinode, $\widetilde\Gamma_{\phi = 0} \approx 2\Gamma$, or double the relaxation rate of the qubit in an open waveguide~\cite{Dorner2002,hoi2015probing,Crowder2020}. The product $\Gamma\tau = 1.097$ tells us the feedback is operating in the non-Markovian regime as the delay time and qubit lifetime are similar. Because $\widetilde\Gamma_{\phi = 0}$ is larger than the other dissipation channels, elastic scattering dominates and the radius of the circle in the complex plane ($|r|$) is $\approx 1$. When $\phi = -0.632\pi$, $\widetilde\Gamma_{\phi = -0.632\pi}$ is much closer to $\gamma_{\rm L} /2 + \gamma'$ and the dip in the reflection coefficient occurs leading to a smaller radius of the circle. Additionally we observe a feedback induced frequency shift of 16 MHz ($\omega_{\rm{0}} = 4.822\,\, \rm{GHz}$ and $\widetilde\omega_{\rm{0}} = 4.806\,\, \rm{GHz}$).

\begin{figure*}
    \centering
    \includegraphics[width=2.1\columnwidth]{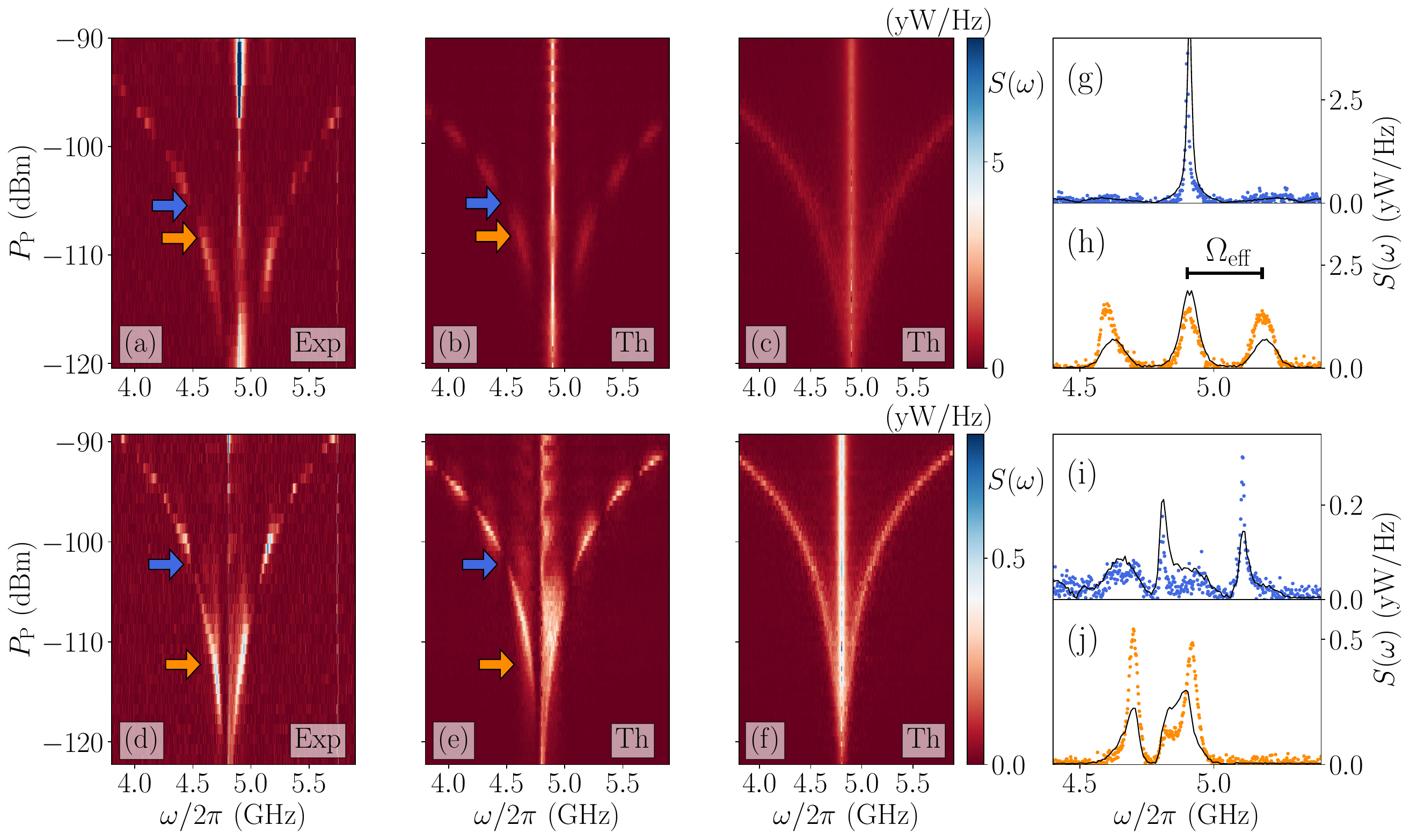}
    \caption{
    %\sh{Can we use the previous one - this is now too big; Franco's comments were suggestions, so let us choose the best ones! Or something in between, now looks too stretched in y (reduce teh y stretching a bit, and to also reduce experimental noise), and this change was nothing to do with referees comments anyway} 
    Resonant fluorescence emission spectra of the qubit as a function of the pump power, $P_{\rm{p}}$ and the detected radiation $\omega$, with $\omega_{\rm{p}}=\widetilde\omega_{0}$.
    %$P_{\rm{p}}$ is the power from RF source. I removed this line since we should define this in the text.
    For (a)-(c) $\phi = 0$ and (d)-(f) $\phi = -0.632 \pi$. In (a) and (d) the experimental results are shown while in (b) and (e) we show the corresponding simulated results. \edit{The bright emission at center peak is due to a multi-level effect (see Sec. V of the SM~\cite{SuppMaterial}). The central peak in (a), (b) and (d), (e) correlate with sideband brightness.} For comparison, the simulated spectral response with a Markovian feedback loop is shown in (c) and (f). Under the strong pump, the qubit energy level split is $\Omega_{\rm{eff}}$, creating three transition frequencies, $\widetilde\omega_{\rm{0}} \pm \Omega_{\rm{eff}}$ and  $\widetilde\omega_{\rm{0}}$. In (g)-(j) we show individual spectra with the experimental data as colored dots and the simulated response as a solid curve. For $\phi = 0$, in (g) $P_{\rm{p}} = -105.4 \, \, \rm{dBm}$ and in (h) $P_{\rm{p}} = -108.4 \, \, \rm{dBm}$. For $\phi = -0.632\pi$, in (i) $P_{\rm{p}} = -102.3 \, \, \rm{dBm}$ and in (j) $P_{\rm{p}} = -112.3 \, \, \rm{dBm}$. The QTDW simulation results use 1000 trajectories per spectrum in (b), (c), (e), and (f) and 10,000 trajectories per simulated spectra in (g)-(j).}
    %\edit{The center peak in (a), (b), (h) and (d), (e) do not match because the excitation pump interfere with the emission.}%~\cite{en2}}
    %Resonant fluorescence emission spectrum as a function of the pump power and detuning of the detected radiation, $\delta\omega_{r}=\omega-\omega_{r}$. $P_q$ is the corresponding power on the qubit. (a) and (b) are measured on the $\phi = 0$, the main feedback feature is that the Mollow sidebands undergo an oscillation in power as the frequency of the sideband passes through points of constructive and destructive interference in the phase of the returning feedback. (b) The linecut pointed by the arrows in (a), red (blue) data shows the constructive (destructive) Mollow triplet. (c) and (d) are measured on the $\phi = 3\pi/4$, where it closes to the node. This makes the sideband structure couples to the TL asymmetric. (d) The linecut pointed by the arrows in (c), the Mollow structure looks very different from the previous work with the feedback delay.
    \label{Mollow}
\end{figure*}

A subtle but interesting non-Markovian effect appears here even in the linear regime. From Eq.~\eqref{ref_coeff_eqn} one expects a symmetric Lorentzian in $|r|$, however in the non-Markovian regime an asymmetry appears which tilts the circle in the complex plane shown in Figs.~\ref{Flux_dp}(e) and (f). The angle of this tilt is $(\widetilde{\omega}_0 - \omega_0)\tau$ and arises because of the feedback induced shift to the qubit frequency. In the Markovian regime ($\tau \sim 0$), all frequencies see a single round trip phase change, $\phi$, while in the non-Markovian regime the round trip phase change for a frequency $\omega'$ is $\phi_{\omega'} = \omega' \tau + \phi_{\rm M} = (\omega' - \omega_0)\tau + \phi$ where the second equivalency we get by writing $\phi_{\rm M}$ in terms of $\phi$. Due to the feedback induced change in the qubit frequency, the round trip phase change that is seen by the qubit emission is $\phi_{\widetilde{\omega}_0} = (\widetilde{\omega}_0 - \omega_0)\tau + \phi$. This readily shows the non-Markovian tilt that is introduced is equivalent to $\phi_{\widetilde{\omega}_0} - \phi$, as shown in Fig.~\ref{Flux_dp}(g), the difference between the round trip phase change of the qubit emission in the non-Markovian and Markovian regimes.

\section*{Resonance Fluorescence Response - Non-linear Regime}

%\sh{Maybe high-field nonlinear regime, to make different to weakly nonlinear? We need main equations in here - Hamiltonian and Effective pump, clarify how we separate coherent and fluctuating parts - then see SM for details. We need a couple of Hamiltonians now in the main text! Too much is buried in the Sup Mat, which can then go unnoticed. And stress again that these models manage to keep observables at the system/quantum level, without the usual approximations ... consistent with the nature and applications of waveguide QED}

When the qubit is pumped by the resonant microwave ($\Delta=0$), its state evolves at an effective Rabi frequency $\Omega_{\rm{eff}}$. When the Rabi frequency of the pumping field is much larger than the natural linewidth of the qubit ($\Omega_{\rm{eff}} \gg \gamma_{\rm{d}}$), the system response includes non-linear dynamics beyond the interference effects of the linear response. In this regime (i.e., without feedback), the energy levels of the qubit become dressed, creating three transition frequencies that appear as distinct spectral components known as the Mollow triplet~\cite{Mollow1969}. These frequencies are $\widetilde\omega_{\rm{0}}$ and $\widetilde\omega_{\rm{0}} \pm \Omega_{\rm{eff}}$. To investigate the effect of feedback on the spectral response of the system, the spectra were recorded as a function of pump power, $P_{\rm{p}}$, by applying the pumping field from a radio frequency (RF) source, resonant with the qubit, to the sample with an effective attenuation, $A$, and sampling the amplified output signal with a spectral analyzer (SA) and normalized to the system gain, $G$. The extracted values are shown in Table \edit{S2 of the SM~\cite{SuppMaterial}}.

Figure~\ref{Mollow} shows the spectral response at two different phase values, with $\phi = 0$ in Figs.~\ref{Mollow}(a)-(c) and $\phi = -0.632 \pi$ in Figs.~\ref{Mollow}(d)-(f). Figures~\ref{Mollow}(c) and (f) show the simulated response with Markovian feedback. In this regime, the feedback is reduced to two phase changes, $\phi$ and $\phi_{\rm{p}}$, which change the effective relaxation rate from the qubit and the effective Rabi frequency respectively. Then, in this regime, as the pump power increases the two sidebands will move further from the central peak following $\widetilde\omega_{\rm{0}} \pm \Omega_{\rm eff}$. Qualitatively, this is the same behavior as the response from a qubit pumped without feedback.

When the feedback is included with a time delay on the order of the qubit lifetime, non-Markovian dynamics appear from the interference between the returning feedback and the qubit output which make qualitative changes to the spectral response. We observed these changes experimentally in Figs.~\ref{Mollow}(a) and (d) which match our simulations in Figs.~\ref{Mollow}(b) and (e), with example slices of the experimentally observed spectra in Figs.~\ref{Mollow}(g)-(j).
For $\phi\,\,=\,\,0$ with $\omega_0/2\pi\,\,=\,\,4.9\,\, \rm{GHz}$, Figs.~\ref{Mollow}(a) and (b), the presence of the Mollow triplet oscillates as the pump power, $P_{\rm{p}}$, increases. 

In contrast to the Markovian regime, where the sidepeaks existed continuously, the macroscopic delay time introduces nodes where the sidepeaks are not observed. These nodes occur when the round trip phase change at that frequency results in destructive interference; $\phi_{\rm node} = (2k-1)\pi$, for $k \in \mathbb{Z}$. This can be written with respect to a phase change that is known, $\phi_{\rm{p}}$, to get
\begin{equation}
    \omega_{\rm node} = \frac{(2k-1)\pi - \phi_{\rm{p}}}{\tau} + \omega_{\rm{p}},
    \label{node_freqs}
\end{equation}
which gives the condition for the nodes seen in Figs.~\ref{Mollow}(a), (b), (d), and (e). This interference condition is symmetric about the central peak when $\phi = 0$ and the pump is on resonance which is seen in the nodes of the red and blue sidebands of the observed spectrum in Figs.~\ref{Mollow}(a) and (b). In Figs.~\ref{Mollow}(g) and (h), we obtained \edit{qualitatively good} agreement between experiment and theory for specific pump powers of $P_{\rm{p}} = -105.4\,\,\rm{dBm}$ and $P_{\rm{p}} = -108.4\,\,\rm{dBm}$. \edit{Differences in the central peak height are due to the notorious difficulty in filtering out the coherent pump field in the experiments.} \edit{In Fig.~\ref{Mollow}(a), the sudden increase of the strong signal is due to the interaction with the higher level with the transmon system, which is an anharmonic oscillator; this level of detail is not included in the theory for simplicity, with more detail discussed in Sec. V of the SM~\cite{SuppMaterial}.}

Next, we tuned $\widetilde\omega_0/2\pi\,\, =\,\,4.806\,\, \rm{GHz}$ and $\phi\,\,=\,\,-0.632 \pi$ in  Figs.~\ref{Mollow}(d) and (e). We excite the qubit at the effective resonance, $\Delta = 0$, with $\phi_{\rm{p}} = -3/4\pi$, due to the difference of $16 \,\,\rm{MHz}$ between the bare and effective resonance frequency. In this case, $\widetilde\Gamma$ is suppressed and the width of the central peak becomes narrower. Moreover, the node structure of the sideband transitions remains when the destructive interference condition is met. However, the red and blue shifted sidebands are suppressed asymmetrically due to the non-zero $\phi$. Again we show spectra at specific pump powers, $P_{\rm{p}} = -102.3\,\,\rm{dBm}$ in Fig.~\ref{Mollow}(i) and $P_{\rm{p}} = -112.3\,\,\rm{dBm}$ in Fig.~\ref{Mollow}(j). \edit{We believe the discrepancy here is primarily driven by differences in the measured dephasing rates in the linear regime and the actual rates in the nonlinear regime. These rates can be dependent on the strength of the driving field~\cite{Cheng2025} and can also change in the presence of the non-Markovian reservoir.}

Additionally, due to the longer qubit lifetime when $\phi = -0.632 \pi$ we can see higher order non-linear dynamics beginning to appear in the spectra. When the feedback is macroscopic, the dressed states of the qubit in the presence of the pumping field are further dressed by the Fabry-P\'{e}rot resonances set up by the feedback loop. These Fabry-P\'{e}rot resonances, which coincide with the antinode frequencies, occur when the round trip phase change at that frequency results in constructive interference;  $\phi_{\rm antinode} = 2 \pi k$~\cite{Crowder2022} for $k \in \mathbb{Z}$. Similar to the nodes, the antinode frequencies can be written as
\begin{equation}
    \omega_{\rm antinode} = \frac{2\pi k - \phi_{\rm{p}}}{\tau} + \omega_{\rm{p}}.
    \label{antinode_freqs}
\end{equation}
This results in multiple new resonances appearing in the spectrum for any $\omega_{\rm antinode}$ between the Mollow sidebands. These are seen in the simulated spectrum of Fig.~\ref{Mollow}(e) and we can begin to see these appearing in the experimental spectrum shown in Fig.~\ref{Mollow}(i) at a pump power of $P_{\rm{p}} =  -102.3\,\,\rm{dBm}$. These results are intrinsic to the feedback system and independent of the specific pumping scheme employed~\cite{Crowder2022}.

\section*{Conclusions}

%In summary, we characterized the qubit and feedback loop in the linear regime through measurement of the reflection coefficient in the presence of a weak probe field. Then explored the effects of the non-Markovian dynamics introduced through the macroscopic delay time of the coherent feedback loop. This is seen through additional nodes appearing the sidebands of the Mollow triplet that appears in the resonance flourescent spectrum. The experimental results produced by the superconducting circuit and transmon qubit platform replicate the simulated results.

We have characterized the superconducting qubit and feedback loop with the waveguide QED system in the linear regime by measuring the reflection coefficient in the presence of a weak pump field. The change in the qubit's frequency and radiative linewidth, as well as the tilt of the reflection coefficient response in the complex plane, are non-Markovian signatures caused by interference with the qubit's own radiation. We then demonstrated the spectrum in the non-linear regime, revealing the Mollow triplet under time-delayed coherent feedback. The distinctive disappearance of the side peaks in resonance fluorescence shows excellent agreement with our QTDW model, suggesting that their formation results from the constructive and destructive interference of emissions with the time-delayed coherent feedback. 

Our work indicates that the atom-mirror system of a superconducting qubit could serve as the simplest platform for engineering a non-Markovian reservoir.
Our results offer a direct approach to accessing the strong coupling regime for the exploration of exotic physics, including long distance collective dipole coupling~\cite{sinha2020non,arranz2021cavitylike}, photon correlations~\cite{Crowder2022} and excitation trapping~\cite{Regidor2021a}. Additionally there are promising applications for quantum technologies such as improving on-demand single photon sources~\cite{Crowder2024}, long distance quantum processing~\cite{breuer2016colloquium} and generating two-dimensional 
photonic cluster states~\cite{Ferreira2024}.
%benefiting from strong coupling and allowing the extension of the atom-mirror distance, making the relaxation rate and delay time comparable.

\acknowledgements

I.-C. H.~acknowledges financial support from City University of Hong Kong through the start-up project 9610569, from the Research Grants Council of Hong Kong (Grant number 11312322), and from the Guangdong Provincial Quantum Science Strategic Initiative (Grant No. GDZX2203001, GDZX2303005, GDZX2403001). This work is supported by the Natural Sciences and Engineering Research of Canada, the National Research Council of Canada, the Canadian Foundation for Innovation, and Queen's University, Canada. S.H. and G.C. acknowledge RIKEN for support. F.N. is supported in part by the Japan Science and Technology Agency (JST) [via the CREST Quantum Frontiers program Grant No. JPMJCR24I2, the Quantum Leap Flagship Program (Q-LEAP), and the Moonshot R\&D Grant Number JPMJMS2061].

\clearpage
\bibliography{PaperBib}

%apsrev4-2.bst 2019-01-14 (MD) hand-edited version of apsrev4-1.bst
%Control: key (0)
%Control: author (8) initials jnrlst
%Control: editor formatted (1) identically to author
%Control: production of article title (0) allowed
%Control: page (0) single
%Control: year (1) truncated
%Control: production of eprint (0) enabled
\begin{thebibliography}{83}%
\makeatletter
\providecommand \@ifxundefined [1]{%
 \@ifx{#1\undefined}
}%
\providecommand \@ifnum [1]{%
 \ifnum #1\expandafter \@firstoftwo
 \else \expandafter \@secondoftwo
 \fi
}%
\providecommand \@ifx [1]{%
 \ifx #1\expandafter \@firstoftwo
 \else \expandafter \@secondoftwo
 \fi
}%
\providecommand \natexlab [1]{#1}%
\providecommand \enquote  [1]{``#1''}%
\providecommand \bibnamefont  [1]{#1}%
\providecommand \bibfnamefont [1]{#1}%
\providecommand \citenamefont [1]{#1}%
\providecommand \href@noop [0]{\@secondoftwo}%
\providecommand \href [0]{\begingroup \@sanitize@url \@href}%
\providecommand \@href[1]{\@@startlink{#1}\@@href}%
\providecommand \@@href[1]{\endgroup#1\@@endlink}%
\providecommand \@sanitize@url [0]{\catcode `\\12\catcode `\$12\catcode `\&12\catcode `\#12\catcode `\^12\catcode `\_12\catcode `\%12\relax}%
\providecommand \@@startlink[1]{}%
\providecommand \@@endlink[0]{}%
\providecommand \url  [0]{\begingroup\@sanitize@url \@url }%
\providecommand \@url [1]{\endgroup\@href {#1}{\urlprefix }}%
\providecommand \urlprefix  [0]{URL }%
\providecommand \Eprint [0]{\href }%
\providecommand \doibase [0]{https://doi.org/}%
\providecommand \selectlanguage [0]{\@gobble}%
\providecommand \bibinfo  [0]{\@secondoftwo}%
\providecommand \bibfield  [0]{\@secondoftwo}%
\providecommand \translation [1]{[#1]}%
\providecommand \BibitemOpen [0]{}%
\providecommand \bibitemStop [0]{}%
\providecommand \bibitemNoStop [0]{.\EOS\space}%
\providecommand \EOS [0]{\spacefactor3000\relax}%
\providecommand \BibitemShut  [1]{\csname bibitem#1\endcsname}%
\let\auto@bib@innerbib\@empty
%</preamble>
\bibitem [{\citenamefont {Kimble}(2008)}]{kimble2008quantum}%
  \BibitemOpen
  \bibfield  {author} {\bibinfo {author} {\bibfnamefont {H.~J.}\ \bibnamefont {Kimble}},\ }\bibfield  {title} {\bibinfo {title} {The quantum internet},\ }\href {https://doi.org/10.1038/nature07127} {\bibfield  {journal} {\bibinfo  {journal} {Nature}\ }\textbf {\bibinfo {volume} {453}},\ \bibinfo {pages} {1023} (\bibinfo {year} {2008})}\BibitemShut {NoStop}%
\bibitem [{\citenamefont {Kimble}\ \emph {et~al.}(1977)\citenamefont {Kimble}, \citenamefont {Dagenais},\ and\ \citenamefont {Mandel}}]{kimble1977photon}%
  \BibitemOpen
  \bibfield  {author} {\bibinfo {author} {\bibfnamefont {H.~J.}\ \bibnamefont {Kimble}}, \bibinfo {author} {\bibfnamefont {M.}~\bibnamefont {Dagenais}},\ and\ \bibinfo {author} {\bibfnamefont {L.}~\bibnamefont {Mandel}},\ }\bibfield  {title} {\bibinfo {title} {Photon antibunching in resonance fluorescence},\ }\href {https://doi.org/10.1103/PhysRevLett.39.691} {\bibfield  {journal} {\bibinfo  {journal} {Physical Review Letters}\ }\textbf {\bibinfo {volume} {39}},\ \bibinfo {pages} {691} (\bibinfo {year} {1977})}\BibitemShut {NoStop}%
\bibitem [{\citenamefont {Eisaman}\ \emph {et~al.}(2011)\citenamefont {Eisaman}, \citenamefont {Fan}, \citenamefont {Migdall},\ and\ \citenamefont {Polyakov}}]{eisaman2011invited}%
  \BibitemOpen
  \bibfield  {author} {\bibinfo {author} {\bibfnamefont {M.~D.}\ \bibnamefont {Eisaman}}, \bibinfo {author} {\bibfnamefont {J.}~\bibnamefont {Fan}}, \bibinfo {author} {\bibfnamefont {A.}~\bibnamefont {Migdall}},\ and\ \bibinfo {author} {\bibfnamefont {S.~V.}\ \bibnamefont {Polyakov}},\ }\bibfield  {title} {\bibinfo {title} {Invited review article: Single-photon sources and detectors},\ }\href {https://doi.org/10.1063/1.3610677} {\bibfield  {journal} {\bibinfo  {journal} {Review of Scientific Instruments}\ }\textbf {\bibinfo {volume} {82}} (\bibinfo {year} {2011})}\BibitemShut {NoStop}%
\bibitem [{\citenamefont {Toyli}\ \emph {et~al.}(2016)\citenamefont {Toyli}, \citenamefont {Eddins}, \citenamefont {Boutin}, \citenamefont {Puri}, \citenamefont {Hover}, \citenamefont {Bolkhovsky}, \citenamefont {Oliver}, \citenamefont {Blais},\ and\ \citenamefont {Siddiqi}}]{toyli2016resonance}%
  \BibitemOpen
  \bibfield  {author} {\bibinfo {author} {\bibfnamefont {D.}~\bibnamefont {Toyli}}, \bibinfo {author} {\bibfnamefont {A.}~\bibnamefont {Eddins}}, \bibinfo {author} {\bibfnamefont {S.}~\bibnamefont {Boutin}}, \bibinfo {author} {\bibfnamefont {S.}~\bibnamefont {Puri}}, \bibinfo {author} {\bibfnamefont {D.}~\bibnamefont {Hover}}, \bibinfo {author} {\bibfnamefont {V.}~\bibnamefont {Bolkhovsky}}, \bibinfo {author} {\bibfnamefont {W.}~\bibnamefont {Oliver}}, \bibinfo {author} {\bibfnamefont {A.}~\bibnamefont {Blais}},\ and\ \bibinfo {author} {\bibfnamefont {I.}~\bibnamefont {Siddiqi}},\ }\bibfield  {title} {\bibinfo {title} {Resonance fluorescence from an artificial atom in squeezed vacuum},\ }\href {https://doi.org/10.1103/PhysRevX.6.031004} {\bibfield  {journal} {\bibinfo  {journal} {Physical Review X}\ }\textbf {\bibinfo {volume} {6}},\ \bibinfo {pages} {031004} (\bibinfo {year} {2016})}\BibitemShut {NoStop}%
\bibitem [{\citenamefont {Lu}\ \emph {et~al.}(2021)\citenamefont {Lu}, \citenamefont {Bengtsson}, \citenamefont {Burnett}, \citenamefont {Wiegand}, \citenamefont {Suri}, \citenamefont {Krantz}, \citenamefont {Roudsari}, \citenamefont {Kockum}, \citenamefont {Gasparinetti}, \citenamefont {Johansson},\ and\ \citenamefont {Delsing}}]{lu2021characterizing}%
  \BibitemOpen
  \bibfield  {author} {\bibinfo {author} {\bibfnamefont {Y.}~\bibnamefont {Lu}}, \bibinfo {author} {\bibfnamefont {A.}~\bibnamefont {Bengtsson}}, \bibinfo {author} {\bibfnamefont {J.~J.}\ \bibnamefont {Burnett}}, \bibinfo {author} {\bibfnamefont {E.}~\bibnamefont {Wiegand}}, \bibinfo {author} {\bibfnamefont {B.}~\bibnamefont {Suri}}, \bibinfo {author} {\bibfnamefont {P.}~\bibnamefont {Krantz}}, \bibinfo {author} {\bibfnamefont {A.~F.}\ \bibnamefont {Roudsari}}, \bibinfo {author} {\bibfnamefont {A.~F.}\ \bibnamefont {Kockum}}, \bibinfo {author} {\bibfnamefont {S.}~\bibnamefont {Gasparinetti}}, \bibinfo {author} {\bibfnamefont {G.}~\bibnamefont {Johansson}},\ and\ \bibinfo {author} {\bibfnamefont {P.}~\bibnamefont {Delsing}},\ }\bibfield  {title} {\bibinfo {title} {Characterizing decoherence rates of a superconducting qubit by direct microwave scattering},\ }\href {https://doi.org/10.1038/s41534-021-00367-5} {\bibfield  {journal} {\bibinfo  {journal} {npj Quantum Information}\ }\textbf {\bibinfo {volume} {7}},\
  \bibinfo {pages} {35} (\bibinfo {year} {2021})}\BibitemShut {NoStop}%
\bibitem [{\citenamefont {Astafiev}\ \emph {et~al.}(2010)\citenamefont {Astafiev}, \citenamefont {Zagoskin}, \citenamefont {Abdumalikov~Jr}, \citenamefont {Pashkin}, \citenamefont {Yamamoto}, \citenamefont {Inomata}, \citenamefont {Nakamura},\ and\ \citenamefont {Tsai}}]{astafiev2010resonance}%
  \BibitemOpen
  \bibfield  {author} {\bibinfo {author} {\bibfnamefont {O.}~\bibnamefont {Astafiev}}, \bibinfo {author} {\bibfnamefont {A.~M.}\ \bibnamefont {Zagoskin}}, \bibinfo {author} {\bibfnamefont {A.}~\bibnamefont {Abdumalikov~Jr}}, \bibinfo {author} {\bibfnamefont {Y.~A.}\ \bibnamefont {Pashkin}}, \bibinfo {author} {\bibfnamefont {T.}~\bibnamefont {Yamamoto}}, \bibinfo {author} {\bibfnamefont {K.}~\bibnamefont {Inomata}}, \bibinfo {author} {\bibfnamefont {Y.}~\bibnamefont {Nakamura}},\ and\ \bibinfo {author} {\bibfnamefont {J.~S.}\ \bibnamefont {Tsai}},\ }\bibfield  {title} {\bibinfo {title} {Resonance fluorescence of a single artificial atom},\ }\href {https://doi.org/10.1126/science.1181918} {\bibfield  {journal} {\bibinfo  {journal} {Science}\ }\textbf {\bibinfo {volume} {327}},\ \bibinfo {pages} {840} (\bibinfo {year} {2010})}\BibitemShut {NoStop}%
\bibitem [{\citenamefont {Kim}\ \emph {et~al.}(2014)\citenamefont {Kim}, \citenamefont {Shen}, \citenamefont {Roy-Choudhury}, \citenamefont {Solomon},\ and\ \citenamefont {Waks}}]{kim2014resonant}%
  \BibitemOpen
  \bibfield  {author} {\bibinfo {author} {\bibfnamefont {H.}~\bibnamefont {Kim}}, \bibinfo {author} {\bibfnamefont {T.~C.}\ \bibnamefont {Shen}}, \bibinfo {author} {\bibfnamefont {K.}~\bibnamefont {Roy-Choudhury}}, \bibinfo {author} {\bibfnamefont {G.~S.}\ \bibnamefont {Solomon}},\ and\ \bibinfo {author} {\bibfnamefont {E.}~\bibnamefont {Waks}},\ }\bibfield  {title} {\bibinfo {title} {Resonant interactions between a {M}ollow triplet sideband and a strongly coupled cavity},\ }\href {https://doi.org/10.1103/PhysRevLett.113.027403} {\bibfield  {journal} {\bibinfo  {journal} {Physical Review Letters}\ }\textbf {\bibinfo {volume} {113}},\ \bibinfo {pages} {027403} (\bibinfo {year} {2014})}\BibitemShut {NoStop}%
\bibitem [{\citenamefont {Kimble}(1976)}]{Kimble1976}%
  \BibitemOpen
  \bibfield  {author} {\bibinfo {author} {\bibfnamefont {H.~J.}\ \bibnamefont {Kimble}},\ }\bibfield  {title} {\bibinfo {title} {Theory of resonance fluorescence},\ }\href {https://doi.org/10.1103/PhysRevA.13.2123} {\bibfield  {journal} {\bibinfo  {journal} {Physical Review A}\ }\textbf {\bibinfo {volume} {13}},\ \bibinfo {pages} {2123} (\bibinfo {year} {1976})}\BibitemShut {NoStop}%
\bibitem [{\citenamefont {Lakowicz}(2006)}]{lakowicz2006principles}%
  \BibitemOpen
  \bibfield  {author} {\bibinfo {author} {\bibfnamefont {J.~R.}\ \bibnamefont {Lakowicz}},\ }\href@noop {} {\emph {\bibinfo {title} {Principles of fluorescence spectroscopy}}},\ \bibinfo {edition} {3rd}\ ed.\ (\bibinfo  {publisher} {Springer},\ \bibinfo {address} {New York, NY},\ \bibinfo {year} {2006})\BibitemShut {NoStop}%
\bibitem [{\citenamefont {Mollow}(1969)}]{Mollow1969}%
  \BibitemOpen
  \bibfield  {author} {\bibinfo {author} {\bibfnamefont {B.~R.}\ \bibnamefont {Mollow}},\ }\bibfield  {title} {\bibinfo {title} {Power {Spectrum} of {Light} {Scattered} by {Two}-{Level} {Systems}},\ }\href {https://doi.org/10.1103/PhysRev.188.1969} {\bibfield  {journal} {\bibinfo  {journal} {Physical Review}\ }\textbf {\bibinfo {volume} {188}},\ \bibinfo {pages} {1969} (\bibinfo {year} {1969})}\BibitemShut {NoStop}%
\bibitem [{\citenamefont {Masters}\ \emph {et~al.}(2023)\citenamefont {Masters}, \citenamefont {Hu}, \citenamefont {Cordier}, \citenamefont {Maron}, \citenamefont {Pache}, \citenamefont {Rauschenbeutel}, \citenamefont {Schemmer},\ and\ \citenamefont {Volz}}]{masters2023simultaneous}%
  \BibitemOpen
  \bibfield  {author} {\bibinfo {author} {\bibfnamefont {L.}~\bibnamefont {Masters}}, \bibinfo {author} {\bibfnamefont {X.-X.}\ \bibnamefont {Hu}}, \bibinfo {author} {\bibfnamefont {M.}~\bibnamefont {Cordier}}, \bibinfo {author} {\bibfnamefont {G.}~\bibnamefont {Maron}}, \bibinfo {author} {\bibfnamefont {L.}~\bibnamefont {Pache}}, \bibinfo {author} {\bibfnamefont {A.}~\bibnamefont {Rauschenbeutel}}, \bibinfo {author} {\bibfnamefont {M.}~\bibnamefont {Schemmer}},\ and\ \bibinfo {author} {\bibfnamefont {J.}~\bibnamefont {Volz}},\ }\bibfield  {title} {\bibinfo {title} {On the simultaneous scattering of two photons by a single two-level atom},\ }\href {https://doi.org/10.1038/s41566-023-01260-7} {\bibfield  {journal} {\bibinfo  {journal} {Nature Photonics}\ }\textbf {\bibinfo {volume} {17}},\ \bibinfo {pages} {972} (\bibinfo {year} {2023})}\BibitemShut {NoStop}%
\bibitem [{\citenamefont {Schuda}\ \emph {et~al.}(1974)\citenamefont {Schuda}, \citenamefont {Stroud~Jr},\ and\ \citenamefont {Hercher}}]{schuda1974observation}%
  \BibitemOpen
  \bibfield  {author} {\bibinfo {author} {\bibfnamefont {F.}~\bibnamefont {Schuda}}, \bibinfo {author} {\bibfnamefont {C.}~\bibnamefont {Stroud~Jr}},\ and\ \bibinfo {author} {\bibfnamefont {M.}~\bibnamefont {Hercher}},\ }\bibfield  {title} {\bibinfo {title} {Observation of the resonant stark effect at optical frequencies},\ }\href {https://doi.org/10.1088/0022-3700/7/7/002} {\bibfield  {journal} {\bibinfo  {journal} {Journal of Physics B: Atomic and Molecular Physics}\ }\textbf {\bibinfo {volume} {7}},\ \bibinfo {pages} {L198} (\bibinfo {year} {1974})}\BibitemShut {NoStop}%
\bibitem [{\citenamefont {Joshi}\ \emph {et~al.}(2023)\citenamefont {Joshi}, \citenamefont {Yang},\ and\ \citenamefont {Mirhosseini}}]{joshi2023resonance}%
  \BibitemOpen
  \bibfield  {author} {\bibinfo {author} {\bibfnamefont {C.}~\bibnamefont {Joshi}}, \bibinfo {author} {\bibfnamefont {F.}~\bibnamefont {Yang}},\ and\ \bibinfo {author} {\bibfnamefont {M.}~\bibnamefont {Mirhosseini}},\ }\bibfield  {title} {\bibinfo {title} {Resonance fluorescence of a chiral artificial atom},\ }\href {https://doi.org/10.1103/PhysRevX.13.021039} {\bibfield  {journal} {\bibinfo  {journal} {Physical Review X}\ }\textbf {\bibinfo {volume} {13}},\ \bibinfo {pages} {021039} (\bibinfo {year} {2023})}\BibitemShut {NoStop}%
\bibitem [{\citenamefont {Redchenko}\ \emph {et~al.}(2023)\citenamefont {Redchenko}, \citenamefont {Poshakinskiy}, \citenamefont {Sett}, \citenamefont {{\v{Z}}emli{\v{c}}ka}, \citenamefont {Poddubny},\ and\ \citenamefont {Fink}}]{redchenko2023tunable}%
  \BibitemOpen
  \bibfield  {author} {\bibinfo {author} {\bibfnamefont {E.~S.}\ \bibnamefont {Redchenko}}, \bibinfo {author} {\bibfnamefont {A.~V.}\ \bibnamefont {Poshakinskiy}}, \bibinfo {author} {\bibfnamefont {R.}~\bibnamefont {Sett}}, \bibinfo {author} {\bibfnamefont {M.}~\bibnamefont {{\v{Z}}emli{\v{c}}ka}}, \bibinfo {author} {\bibfnamefont {A.~N.}\ \bibnamefont {Poddubny}},\ and\ \bibinfo {author} {\bibfnamefont {J.~M.}\ \bibnamefont {Fink}},\ }\bibfield  {title} {\bibinfo {title} {Tunable directional photon scattering from a pair of superconducting qubits},\ }\href {https://doi.org/10.5281/zenodo.7858567} {\bibfield  {journal} {\bibinfo  {journal} {Nature Communications}\ }\textbf {\bibinfo {volume} {14}},\ \bibinfo {pages} {2998} (\bibinfo {year} {2023})}\BibitemShut {NoStop}%
\bibitem [{\citenamefont {Nick~Vamivakas}\ \emph {et~al.}(2009)\citenamefont {Nick~Vamivakas}, \citenamefont {Zhao}, \citenamefont {Lu},\ and\ \citenamefont {Atatüre}}]{nick2009spin}%
  \BibitemOpen
  \bibfield  {author} {\bibinfo {author} {\bibfnamefont {A.}~\bibnamefont {Nick~Vamivakas}}, \bibinfo {author} {\bibfnamefont {Y.}~\bibnamefont {Zhao}}, \bibinfo {author} {\bibfnamefont {C.-Y.}\ \bibnamefont {Lu}},\ and\ \bibinfo {author} {\bibfnamefont {M.}~\bibnamefont {Atatüre}},\ }\bibfield  {title} {\bibinfo {title} {Spin-resolved quantum-dot resonance fluorescence},\ }\href {https://doi.org/10.1038/nphys1182} {\bibfield  {journal} {\bibinfo  {journal} {Nature Physics}\ }\textbf {\bibinfo {volume} {5}},\ \bibinfo {pages} {198} (\bibinfo {year} {2009})}\BibitemShut {NoStop}%
\bibitem [{\citenamefont {Ulhaq}\ \emph {et~al.}(2013)\citenamefont {Ulhaq}, \citenamefont {Weiler}, \citenamefont {Roy}, \citenamefont {Ulrich}, \citenamefont {Jetter}, \citenamefont {Hughes},\ and\ \citenamefont {Michler}}]{Ulhaq2013}%
  \BibitemOpen
  \bibfield  {author} {\bibinfo {author} {\bibfnamefont {A.}~\bibnamefont {Ulhaq}}, \bibinfo {author} {\bibfnamefont {S.}~\bibnamefont {Weiler}}, \bibinfo {author} {\bibfnamefont {C.}~\bibnamefont {Roy}}, \bibinfo {author} {\bibfnamefont {S.~M.}\ \bibnamefont {Ulrich}}, \bibinfo {author} {\bibfnamefont {M.}~\bibnamefont {Jetter}}, \bibinfo {author} {\bibfnamefont {S.}~\bibnamefont {Hughes}},\ and\ \bibinfo {author} {\bibfnamefont {P.}~\bibnamefont {Michler}},\ }\bibfield  {title} {\bibinfo {title} {Detuning-dependent mollow triplet of a coherently-driven single quantum dot},\ }\href {https://doi.org/10.1364/oe.21.004382} {\bibfield  {journal} {\bibinfo  {journal} {Optics Express}\ }\textbf {\bibinfo {volume} {21}},\ \bibinfo {pages} {4382} (\bibinfo {year} {2013})}\BibitemShut {NoStop}%
\bibitem [{\citenamefont {Gasparinetti}\ \emph {et~al.}(2019)\citenamefont {Gasparinetti}, \citenamefont {Besse}, \citenamefont {Pechal}, \citenamefont {Buijs}, \citenamefont {Eichler}, \citenamefont {Carmichael},\ and\ \citenamefont {Wallraff}}]{gasparinetti2019two}%
  \BibitemOpen
  \bibfield  {author} {\bibinfo {author} {\bibfnamefont {S.}~\bibnamefont {Gasparinetti}}, \bibinfo {author} {\bibfnamefont {J.-C.}\ \bibnamefont {Besse}}, \bibinfo {author} {\bibfnamefont {M.}~\bibnamefont {Pechal}}, \bibinfo {author} {\bibfnamefont {R.~D.}\ \bibnamefont {Buijs}}, \bibinfo {author} {\bibfnamefont {C.}~\bibnamefont {Eichler}}, \bibinfo {author} {\bibfnamefont {H.~J.}\ \bibnamefont {Carmichael}},\ and\ \bibinfo {author} {\bibfnamefont {A.}~\bibnamefont {Wallraff}},\ }\bibfield  {title} {\bibinfo {title} {Two-photon resonance fluorescence of a ladder-type atomic system},\ }\href {https://doi.org/10.1103/PhysRevA.100.033802} {\bibfield  {journal} {\bibinfo  {journal} {Physical Review A}\ }\textbf {\bibinfo {volume} {100}},\ \bibinfo {pages} {033802} (\bibinfo {year} {2019})}\BibitemShut {NoStop}%
\bibitem [{\citenamefont {Ruan}\ \emph {et~al.}(2024)\citenamefont {Ruan}, \citenamefont {Wang}, \citenamefont {He}, \citenamefont {Song}, \citenamefont {Li}, \citenamefont {Zhao}, \citenamefont {Kuang}, \citenamefont {Tsai}, \citenamefont {Zou}, \citenamefont {Zhang} \emph {et~al.}}]{ruan2024dynamics}%
  \BibitemOpen
  \bibfield  {author} {\bibinfo {author} {\bibfnamefont {X.}~\bibnamefont {Ruan}}, \bibinfo {author} {\bibfnamefont {J.-H.}\ \bibnamefont {Wang}}, \bibinfo {author} {\bibfnamefont {D.}~\bibnamefont {He}}, \bibinfo {author} {\bibfnamefont {P.}~\bibnamefont {Song}}, \bibinfo {author} {\bibfnamefont {S.}~\bibnamefont {Li}}, \bibinfo {author} {\bibfnamefont {Q.}~\bibnamefont {Zhao}}, \bibinfo {author} {\bibfnamefont {L.}~\bibnamefont {Kuang}}, \bibinfo {author} {\bibfnamefont {J.-S.}\ \bibnamefont {Tsai}}, \bibinfo {author} {\bibfnamefont {C.-L.}\ \bibnamefont {Zou}}, \bibinfo {author} {\bibfnamefont {J.}~\bibnamefont {Zhang}}, \emph {et~al.},\ }\bibfield  {title} {\bibinfo {title} {Dynamics and resonance fluorescence from a superconducting artificial atom doubly driven by quantized and classical fields},\ }\href {https://doi.org/10.1103/PhysRevResearch.6.033064} {\bibfield  {journal} {\bibinfo  {journal} {Physical Review Research}\ }\textbf {\bibinfo {volume} {6}},\ \bibinfo {pages} {033064} (\bibinfo {year}
  {2024})}\BibitemShut {NoStop}%
\bibitem [{\citenamefont {Boos}\ \emph {et~al.}(2024)\citenamefont {Boos}, \citenamefont {Kim}, \citenamefont {Bracht}, \citenamefont {Sbresny}, \citenamefont {Kaspari}, \citenamefont {Cygorek}, \citenamefont {Riedl}, \citenamefont {Bopp}, \citenamefont {Rauhaus}, \citenamefont {Calcagno} \emph {et~al.}}]{boos2024signatures}%
  \BibitemOpen
  \bibfield  {author} {\bibinfo {author} {\bibfnamefont {K.}~\bibnamefont {Boos}}, \bibinfo {author} {\bibfnamefont {S.~K.}\ \bibnamefont {Kim}}, \bibinfo {author} {\bibfnamefont {T.}~\bibnamefont {Bracht}}, \bibinfo {author} {\bibfnamefont {F.}~\bibnamefont {Sbresny}}, \bibinfo {author} {\bibfnamefont {J.~M.}\ \bibnamefont {Kaspari}}, \bibinfo {author} {\bibfnamefont {M.}~\bibnamefont {Cygorek}}, \bibinfo {author} {\bibfnamefont {H.}~\bibnamefont {Riedl}}, \bibinfo {author} {\bibfnamefont {F.~W.}\ \bibnamefont {Bopp}}, \bibinfo {author} {\bibfnamefont {W.}~\bibnamefont {Rauhaus}}, \bibinfo {author} {\bibfnamefont {C.}~\bibnamefont {Calcagno}}, \emph {et~al.},\ }\bibfield  {title} {\bibinfo {title} {Signatures of dynamically dressed states},\ }\href {https://doi.org/10.1103/PhysRevLett.132.053602} {\bibfield  {journal} {\bibinfo  {journal} {Physical Review Letters}\ }\textbf {\bibinfo {volume} {132}},\ \bibinfo {pages} {053602} (\bibinfo {year} {2024})}\BibitemShut {NoStop}%
\bibitem [{\citenamefont {Liu}\ \emph {et~al.}(2024)\citenamefont {Liu}, \citenamefont {Gustin}, \citenamefont {Liu}, \citenamefont {Li}, \citenamefont {Yu}, \citenamefont {Ni}, \citenamefont {Niu}, \citenamefont {Hughes}, \citenamefont {Wang},\ and\ \citenamefont {Liu}}]{liu2024dynamic}%
  \BibitemOpen
  \bibfield  {author} {\bibinfo {author} {\bibfnamefont {S.}~\bibnamefont {Liu}}, \bibinfo {author} {\bibfnamefont {C.}~\bibnamefont {Gustin}}, \bibinfo {author} {\bibfnamefont {H.}~\bibnamefont {Liu}}, \bibinfo {author} {\bibfnamefont {X.}~\bibnamefont {Li}}, \bibinfo {author} {\bibfnamefont {Y.}~\bibnamefont {Yu}}, \bibinfo {author} {\bibfnamefont {H.}~\bibnamefont {Ni}}, \bibinfo {author} {\bibfnamefont {Z.}~\bibnamefont {Niu}}, \bibinfo {author} {\bibfnamefont {S.}~\bibnamefont {Hughes}}, \bibinfo {author} {\bibfnamefont {X.}~\bibnamefont {Wang}},\ and\ \bibinfo {author} {\bibfnamefont {J.}~\bibnamefont {Liu}},\ }\bibfield  {title} {\bibinfo {title} {Dynamic resonance fluorescence in solid-state cavity quantum electrodynamics},\ }\href {https://doi.org/10.6084/m9.figshare.24501241} {\bibfield  {journal} {\bibinfo  {journal} {Nature Photonics}\ }\textbf {\bibinfo {volume} {18}},\ \bibinfo {pages} {318} (\bibinfo {year} {2024})}\BibitemShut {NoStop}%
\bibitem [{\citenamefont {L{\'o}pez~Carre{\~n}o}\ \emph {et~al.}(2024)\citenamefont {L{\'o}pez~Carre{\~n}o}, \citenamefont {Berm{\'u}dez~Feijoo},\ and\ \citenamefont {Stobi{\'n}ska}}]{lopez2024entanglement}%
  \BibitemOpen
  \bibfield  {author} {\bibinfo {author} {\bibfnamefont {J.~C.}\ \bibnamefont {L{\'o}pez~Carre{\~n}o}}, \bibinfo {author} {\bibfnamefont {S.}~\bibnamefont {Berm{\'u}dez~Feijoo}},\ and\ \bibinfo {author} {\bibfnamefont {M.}~\bibnamefont {Stobi{\'n}ska}},\ }\bibfield  {title} {\bibinfo {title} {Entanglement in resonance fluorescence},\ }\href {https://doi.org/10.7910/DVN/TLBG9X} {\bibfield  {journal} {\bibinfo  {journal} {npj Nanophotonics}\ }\textbf {\bibinfo {volume} {1}},\ \bibinfo {pages} {3} (\bibinfo {year} {2024})}\BibitemShut {NoStop}%
\bibitem [{\citenamefont {Blais}\ \emph {et~al.}(2021)\citenamefont {Blais}, \citenamefont {Grimsmo}, \citenamefont {Girvin},\ and\ \citenamefont {Wallraff}}]{blais2021circuit}%
  \BibitemOpen
  \bibfield  {author} {\bibinfo {author} {\bibfnamefont {A.}~\bibnamefont {Blais}}, \bibinfo {author} {\bibfnamefont {A.~L.}\ \bibnamefont {Grimsmo}}, \bibinfo {author} {\bibfnamefont {S.~M.}\ \bibnamefont {Girvin}},\ and\ \bibinfo {author} {\bibfnamefont {A.}~\bibnamefont {Wallraff}},\ }\bibfield  {title} {\bibinfo {title} {Circuit quantum electrodynamics},\ }\href {https://doi.org/10.1103/RevModPhys.93.025005} {\bibfield  {journal} {\bibinfo  {journal} {Reviews of Modern Physics}\ }\textbf {\bibinfo {volume} {93}},\ \bibinfo {pages} {025005} (\bibinfo {year} {2021})}\BibitemShut {NoStop}%
\bibitem [{\citenamefont {Kockum}\ and\ \citenamefont {Nori}(2019)}]{kockum2019quantum}%
  \BibitemOpen
  \bibfield  {author} {\bibinfo {author} {\bibfnamefont {A.~F.}\ \bibnamefont {Kockum}}\ and\ \bibinfo {author} {\bibfnamefont {F.}~\bibnamefont {Nori}},\ }\bibfield  {title} {\bibinfo {title} {Quantum bits with {J}osephson junctions},\ }in\ \href@noop {} {\emph {\bibinfo {booktitle} {Fundamentals and Frontiers of the Josephson Effect}}}\ (\bibinfo  {publisher} {Springer},\ \bibinfo {year} {2019})\ pp.\ \bibinfo {pages} {703--741}\BibitemShut {NoStop}%
\bibitem [{\citenamefont {Gu}\ \emph {et~al.}(2017)\citenamefont {Gu}, \citenamefont {Kockum}, \citenamefont {Miranowicz}, \citenamefont {Liu},\ and\ \citenamefont {Nori}}]{gu2017microwave}%
  \BibitemOpen
  \bibfield  {author} {\bibinfo {author} {\bibfnamefont {X.}~\bibnamefont {Gu}}, \bibinfo {author} {\bibfnamefont {A.~F.}\ \bibnamefont {Kockum}}, \bibinfo {author} {\bibfnamefont {A.}~\bibnamefont {Miranowicz}}, \bibinfo {author} {\bibfnamefont {Y.-x.}\ \bibnamefont {Liu}},\ and\ \bibinfo {author} {\bibfnamefont {F.}~\bibnamefont {Nori}},\ }\bibfield  {title} {\bibinfo {title} {Microwave photonics with superconducting quantum circuits},\ }\href {https://doi.org/10.1016/j.physrep.2017.10.002} {\bibfield  {journal} {\bibinfo  {journal} {Physics Reports}\ }\textbf {\bibinfo {volume} {718-719}},\ \bibinfo {pages} {1} (\bibinfo {year} {2017})}\BibitemShut {NoStop}%
\bibitem [{\citenamefont {Kockum}\ \emph {et~al.}(2018)\citenamefont {Kockum}, \citenamefont {Johansson},\ and\ \citenamefont {Nori}}]{kockum2018decoherence}%
  \BibitemOpen
  \bibfield  {author} {\bibinfo {author} {\bibfnamefont {A.~F.}\ \bibnamefont {Kockum}}, \bibinfo {author} {\bibfnamefont {G.}~\bibnamefont {Johansson}},\ and\ \bibinfo {author} {\bibfnamefont {F.}~\bibnamefont {Nori}},\ }\bibfield  {title} {\bibinfo {title} {Decoherence-free interaction between giant atoms in waveguide quantum electrodynamics},\ }\href@noop {} {\bibfield  {journal} {\bibinfo  {journal} {Physical review letters}\ }\textbf {\bibinfo {volume} {120}},\ \bibinfo {pages} {140404} (\bibinfo {year} {2018})}\BibitemShut {NoStop}%
\bibitem [{\citenamefont {Kannan}\ \emph {et~al.}(2020)\citenamefont {Kannan}, \citenamefont {Ruckriegel}, \citenamefont {Campbell}, \citenamefont {Frisk~Kockum}, \citenamefont {Braum{\"u}ller}, \citenamefont {Kim}, \citenamefont {Kjaergaard}, \citenamefont {Krantz}, \citenamefont {Melville}, \citenamefont {Niedzielski} \emph {et~al.}}]{kannan2020waveguide}%
  \BibitemOpen
  \bibfield  {author} {\bibinfo {author} {\bibfnamefont {B.}~\bibnamefont {Kannan}}, \bibinfo {author} {\bibfnamefont {M.~J.}\ \bibnamefont {Ruckriegel}}, \bibinfo {author} {\bibfnamefont {D.~L.}\ \bibnamefont {Campbell}}, \bibinfo {author} {\bibfnamefont {A.}~\bibnamefont {Frisk~Kockum}}, \bibinfo {author} {\bibfnamefont {J.}~\bibnamefont {Braum{\"u}ller}}, \bibinfo {author} {\bibfnamefont {D.~K.}\ \bibnamefont {Kim}}, \bibinfo {author} {\bibfnamefont {M.}~\bibnamefont {Kjaergaard}}, \bibinfo {author} {\bibfnamefont {P.}~\bibnamefont {Krantz}}, \bibinfo {author} {\bibfnamefont {A.}~\bibnamefont {Melville}}, \bibinfo {author} {\bibfnamefont {B.~M.}\ \bibnamefont {Niedzielski}}, \emph {et~al.},\ }\bibfield  {title} {\bibinfo {title} {Waveguide quantum electrodynamics with superconducting artificial giant atoms},\ }\href {https://doi.org/10.1038/s41586-020-2529-9} {\bibfield  {journal} {\bibinfo  {journal} {Nature}\ }\textbf {\bibinfo {volume} {583}},\ \bibinfo {pages} {775} (\bibinfo {year}
  {2020})}\BibitemShut {NoStop}%
\bibitem [{\citenamefont {Liul}\ \emph {et~al.}(2023)\citenamefont {Liul}, \citenamefont {Chien}, \citenamefont {Chen}, \citenamefont {Wen}, \citenamefont {Chen}, \citenamefont {Lin}, \citenamefont {Shevchenko}, \citenamefont {Nori},\ and\ \citenamefont {Hoi}}]{liul2023coherent}%
  \BibitemOpen
  \bibfield  {author} {\bibinfo {author} {\bibfnamefont {M.}~\bibnamefont {Liul}}, \bibinfo {author} {\bibfnamefont {C.-H.}\ \bibnamefont {Chien}}, \bibinfo {author} {\bibfnamefont {C.-Y.}\ \bibnamefont {Chen}}, \bibinfo {author} {\bibfnamefont {P.}~\bibnamefont {Wen}}, \bibinfo {author} {\bibfnamefont {J.}~\bibnamefont {Chen}}, \bibinfo {author} {\bibfnamefont {Y.-H.}\ \bibnamefont {Lin}}, \bibinfo {author} {\bibfnamefont {S.}~\bibnamefont {Shevchenko}}, \bibinfo {author} {\bibfnamefont {F.}~\bibnamefont {Nori}},\ and\ \bibinfo {author} {\bibfnamefont {I.-C.}\ \bibnamefont {Hoi}},\ }\bibfield  {title} {\bibinfo {title} {Coherent dynamics of a photon-dressed qubit},\ }\href@noop {} {\bibfield  {journal} {\bibinfo  {journal} {Physical Review B}\ }\textbf {\bibinfo {volume} {107}},\ \bibinfo {pages} {195441} (\bibinfo {year} {2023})}\BibitemShut {NoStop}%
\bibitem [{\citenamefont {Kurpiers}\ \emph {et~al.}(2018)\citenamefont {Kurpiers}, \citenamefont {Magnard}, \citenamefont {Walter}, \citenamefont {Royer}, \citenamefont {Pechal}, \citenamefont {Heinsoo}, \citenamefont {Salath{\'e}}, \citenamefont {Akin}, \citenamefont {Storz}, \citenamefont {Besse} \emph {et~al.}}]{kurpiers2018deterministic}%
  \BibitemOpen
  \bibfield  {author} {\bibinfo {author} {\bibfnamefont {P.}~\bibnamefont {Kurpiers}}, \bibinfo {author} {\bibfnamefont {P.}~\bibnamefont {Magnard}}, \bibinfo {author} {\bibfnamefont {T.}~\bibnamefont {Walter}}, \bibinfo {author} {\bibfnamefont {B.}~\bibnamefont {Royer}}, \bibinfo {author} {\bibfnamefont {M.}~\bibnamefont {Pechal}}, \bibinfo {author} {\bibfnamefont {J.}~\bibnamefont {Heinsoo}}, \bibinfo {author} {\bibfnamefont {Y.}~\bibnamefont {Salath{\'e}}}, \bibinfo {author} {\bibfnamefont {A.}~\bibnamefont {Akin}}, \bibinfo {author} {\bibfnamefont {S.}~\bibnamefont {Storz}}, \bibinfo {author} {\bibfnamefont {J.-C.}\ \bibnamefont {Besse}}, \emph {et~al.},\ }\bibfield  {title} {\bibinfo {title} {Deterministic quantum state transfer and remote entanglement using microwave photons},\ }\href {https://doi.org/10.1038/s41586-018-0195-y} {\bibfield  {journal} {\bibinfo  {journal} {Nature}\ }\textbf {\bibinfo {volume} {558}},\ \bibinfo {pages} {264} (\bibinfo {year} {2018})}\BibitemShut {NoStop}%
\bibitem [{\citenamefont {Hoi}\ \emph {et~al.}(2011)\citenamefont {Hoi}, \citenamefont {Wilson}, \citenamefont {Johansson}, \citenamefont {Palomaki}, \citenamefont {Peropadre},\ and\ \citenamefont {Delsing}}]{hoi2011demonstration}%
  \BibitemOpen
  \bibfield  {author} {\bibinfo {author} {\bibfnamefont {I.-C.}\ \bibnamefont {Hoi}}, \bibinfo {author} {\bibfnamefont {C.}~\bibnamefont {Wilson}}, \bibinfo {author} {\bibfnamefont {G.}~\bibnamefont {Johansson}}, \bibinfo {author} {\bibfnamefont {T.}~\bibnamefont {Palomaki}}, \bibinfo {author} {\bibfnamefont {B.}~\bibnamefont {Peropadre}},\ and\ \bibinfo {author} {\bibfnamefont {P.}~\bibnamefont {Delsing}},\ }\bibfield  {title} {\bibinfo {title} {Demonstration of a single-photon router in the microwave regime},\ }\href {https://doi.org/10.1103/PhysRevLett.107.073601} {\bibfield  {journal} {\bibinfo  {journal} {Physical Review Letters}\ }\textbf {\bibinfo {volume} {107}},\ \bibinfo {pages} {073601} (\bibinfo {year} {2011})}\BibitemShut {NoStop}%
\bibitem [{\citenamefont {Hoi}\ \emph {et~al.}(2012)\citenamefont {Hoi}, \citenamefont {Palomaki}, \citenamefont {Lindkvist}, \citenamefont {Johansson}, \citenamefont {Delsing},\ and\ \citenamefont {Wilson}}]{hoi2012generation}%
  \BibitemOpen
  \bibfield  {author} {\bibinfo {author} {\bibfnamefont {I.-C.}\ \bibnamefont {Hoi}}, \bibinfo {author} {\bibfnamefont {T.}~\bibnamefont {Palomaki}}, \bibinfo {author} {\bibfnamefont {J.}~\bibnamefont {Lindkvist}}, \bibinfo {author} {\bibfnamefont {G.}~\bibnamefont {Johansson}}, \bibinfo {author} {\bibfnamefont {P.}~\bibnamefont {Delsing}},\ and\ \bibinfo {author} {\bibfnamefont {C.}~\bibnamefont {Wilson}},\ }\bibfield  {title} {\bibinfo {title} {Generation of nonclassical microwave states using an artificial atom in 1d open space},\ }\href {https://doi.org/10.1103/PhysRevLett.108.263601} {\bibfield  {journal} {\bibinfo  {journal} {Physical Review Letters}\ }\textbf {\bibinfo {volume} {108}},\ \bibinfo {pages} {263601} (\bibinfo {year} {2012})}\BibitemShut {NoStop}%
\bibitem [{\citenamefont {Hoi}\ \emph {et~al.}(2013)\citenamefont {Hoi}, \citenamefont {Kockum}, \citenamefont {Palomaki}, \citenamefont {Stace}, \citenamefont {Fan}, \citenamefont {Tornberg}, \citenamefont {Sathyamoorthy}, \citenamefont {Johansson}, \citenamefont {Delsing},\ and\ \citenamefont {Wilson}}]{hoi2013giant}%
  \BibitemOpen
  \bibfield  {author} {\bibinfo {author} {\bibfnamefont {I.-C.}\ \bibnamefont {Hoi}}, \bibinfo {author} {\bibfnamefont {A.~F.}\ \bibnamefont {Kockum}}, \bibinfo {author} {\bibfnamefont {T.}~\bibnamefont {Palomaki}}, \bibinfo {author} {\bibfnamefont {T.~M.}\ \bibnamefont {Stace}}, \bibinfo {author} {\bibfnamefont {B.}~\bibnamefont {Fan}}, \bibinfo {author} {\bibfnamefont {L.}~\bibnamefont {Tornberg}}, \bibinfo {author} {\bibfnamefont {S.~R.}\ \bibnamefont {Sathyamoorthy}}, \bibinfo {author} {\bibfnamefont {G.}~\bibnamefont {Johansson}}, \bibinfo {author} {\bibfnamefont {P.}~\bibnamefont {Delsing}},\ and\ \bibinfo {author} {\bibfnamefont {C.}~\bibnamefont {Wilson}},\ }\bibfield  {title} {\bibinfo {title} {Giant cross--kerr effect for propagating microwaves induced by an artificial atom},\ }\href {https://doi.org/10.1103/PhysRevLett.111.053601} {\bibfield  {journal} {\bibinfo  {journal} {Physical Review Letters}\ }\textbf {\bibinfo {volume} {111}},\ \bibinfo {pages} {053601} (\bibinfo {year} {2013})}\BibitemShut
  {NoStop}%
\bibitem [{\citenamefont {Hoi}\ \emph {et~al.}(2015)\citenamefont {Hoi}, \citenamefont {Kockum}, \citenamefont {Tornberg}, \citenamefont {Pourkabirian}, \citenamefont {Johansson}, \citenamefont {Delsing},\ and\ \citenamefont {Wilson}}]{hoi2015probing}%
  \BibitemOpen
  \bibfield  {author} {\bibinfo {author} {\bibfnamefont {I.-C.}\ \bibnamefont {Hoi}}, \bibinfo {author} {\bibfnamefont {A.~F.}\ \bibnamefont {Kockum}}, \bibinfo {author} {\bibfnamefont {L.}~\bibnamefont {Tornberg}}, \bibinfo {author} {\bibfnamefont {A.}~\bibnamefont {Pourkabirian}}, \bibinfo {author} {\bibfnamefont {G.}~\bibnamefont {Johansson}}, \bibinfo {author} {\bibfnamefont {P.}~\bibnamefont {Delsing}},\ and\ \bibinfo {author} {\bibfnamefont {C.}~\bibnamefont {Wilson}},\ }\bibfield  {title} {\bibinfo {title} {Probing the quantum vacuum with an artificial atom in front of a mirror},\ }\href {https://doi.org/10.1038/nphys3484} {\bibfield  {journal} {\bibinfo  {journal} {Nature Physics}\ }\textbf {\bibinfo {volume} {11}},\ \bibinfo {pages} {1045} (\bibinfo {year} {2015})}\BibitemShut {NoStop}%
\bibitem [{\citenamefont {Wen}\ \emph {et~al.}(2018)\citenamefont {Wen}, \citenamefont {Kockum}, \citenamefont {Ian}, \citenamefont {Chen}, \citenamefont {Nori},\ and\ \citenamefont {Hoi}}]{wen2018reflective}%
  \BibitemOpen
  \bibfield  {author} {\bibinfo {author} {\bibfnamefont {P.}~\bibnamefont {Wen}}, \bibinfo {author} {\bibfnamefont {A.}~\bibnamefont {Kockum}}, \bibinfo {author} {\bibfnamefont {H.}~\bibnamefont {Ian}}, \bibinfo {author} {\bibfnamefont {J.}~\bibnamefont {Chen}}, \bibinfo {author} {\bibfnamefont {F.}~\bibnamefont {Nori}},\ and\ \bibinfo {author} {\bibfnamefont {I.-C.}\ \bibnamefont {Hoi}},\ }\bibfield  {title} {\bibinfo {title} {Reflective amplification without population inversion from a strongly driven superconducting qubit},\ }\href {https://doi.org/10.1103/PhysRevLett.120.063603} {\bibfield  {journal} {\bibinfo  {journal} {Physical Review Letters}\ }\textbf {\bibinfo {volume} {120}},\ \bibinfo {pages} {063603} (\bibinfo {year} {2018})}\BibitemShut {NoStop}%
\bibitem [{\citenamefont {Wen}\ \emph {et~al.}(2019)\citenamefont {Wen}, \citenamefont {Lin}, \citenamefont {Kockum}, \citenamefont {Suri}, \citenamefont {Ian}, \citenamefont {Chen}, \citenamefont {Mao}, \citenamefont {Chiu}, \citenamefont {Delsing}, \citenamefont {Nori} \emph {et~al.}}]{wen2019large}%
  \BibitemOpen
  \bibfield  {author} {\bibinfo {author} {\bibfnamefont {P.}~\bibnamefont {Wen}}, \bibinfo {author} {\bibfnamefont {K.-T.}\ \bibnamefont {Lin}}, \bibinfo {author} {\bibfnamefont {A.}~\bibnamefont {Kockum}}, \bibinfo {author} {\bibfnamefont {B.}~\bibnamefont {Suri}}, \bibinfo {author} {\bibfnamefont {H.}~\bibnamefont {Ian}}, \bibinfo {author} {\bibfnamefont {J.}~\bibnamefont {Chen}}, \bibinfo {author} {\bibfnamefont {S.}~\bibnamefont {Mao}}, \bibinfo {author} {\bibfnamefont {C.}~\bibnamefont {Chiu}}, \bibinfo {author} {\bibfnamefont {P.}~\bibnamefont {Delsing}}, \bibinfo {author} {\bibfnamefont {F.}~\bibnamefont {Nori}}, \emph {et~al.},\ }\bibfield  {title} {\bibinfo {title} {Large collective lamb shift of two distant superconducting artificial atoms},\ }\href {https://doi.org/10.1103/PhysRevLett.123.233602} {\bibfield  {journal} {\bibinfo  {journal} {Physical Review Letters}\ }\textbf {\bibinfo {volume} {123}},\ \bibinfo {pages} {233602} (\bibinfo {year} {2019})}\BibitemShut {NoStop}%
\bibitem [{\citenamefont {Lin}\ \emph {et~al.}(2019)\citenamefont {Lin}, \citenamefont {Hsu}, \citenamefont {Lee}, \citenamefont {Hoi},\ and\ \citenamefont {Lin}}]{lin2019scalable}%
  \BibitemOpen
  \bibfield  {author} {\bibinfo {author} {\bibfnamefont {K.-T.}\ \bibnamefont {Lin}}, \bibinfo {author} {\bibfnamefont {T.}~\bibnamefont {Hsu}}, \bibinfo {author} {\bibfnamefont {C.-Y.}\ \bibnamefont {Lee}}, \bibinfo {author} {\bibfnamefont {I.-C.}\ \bibnamefont {Hoi}},\ and\ \bibinfo {author} {\bibfnamefont {G.-D.}\ \bibnamefont {Lin}},\ }\bibfield  {title} {\bibinfo {title} {Scalable collective lamb shift of a 1d superconducting qubit array in front of a mirror},\ }\href@noop {} {\bibfield  {journal} {\bibinfo  {journal} {Scientific reports}\ }\textbf {\bibinfo {volume} {9}},\ \bibinfo {pages} {19175} (\bibinfo {year} {2019})}\BibitemShut {NoStop}%
\bibitem [{\citenamefont {Van~Loo}\ \emph {et~al.}(2013)\citenamefont {Van~Loo}, \citenamefont {Fedorov}, \citenamefont {Lalumiere}, \citenamefont {Sanders}, \citenamefont {Blais},\ and\ \citenamefont {Wallraff}}]{van2013photon}%
  \BibitemOpen
  \bibfield  {author} {\bibinfo {author} {\bibfnamefont {A.~F.}\ \bibnamefont {Van~Loo}}, \bibinfo {author} {\bibfnamefont {A.}~\bibnamefont {Fedorov}}, \bibinfo {author} {\bibfnamefont {K.}~\bibnamefont {Lalumiere}}, \bibinfo {author} {\bibfnamefont {B.~C.}\ \bibnamefont {Sanders}}, \bibinfo {author} {\bibfnamefont {A.}~\bibnamefont {Blais}},\ and\ \bibinfo {author} {\bibfnamefont {A.}~\bibnamefont {Wallraff}},\ }\bibfield  {title} {\bibinfo {title} {Photon-mediated interactions between distant artificial atoms},\ }\href {https://doi.org/10.1126/science.1244324} {\bibfield  {journal} {\bibinfo  {journal} {Science}\ }\textbf {\bibinfo {volume} {342}},\ \bibinfo {pages} {1494} (\bibinfo {year} {2013})}\BibitemShut {NoStop}%
\bibitem [{\citenamefont {Kannan}\ \emph {et~al.}(2023)\citenamefont {Kannan}, \citenamefont {Almanakly}, \citenamefont {Sung}, \citenamefont {Di~Paolo}, \citenamefont {Rower}, \citenamefont {Braum{\"u}ller}, \citenamefont {Melville}, \citenamefont {Niedzielski}, \citenamefont {Karamlou}, \citenamefont {Serniak} \emph {et~al.}}]{kannan2023demand}%
  \BibitemOpen
  \bibfield  {author} {\bibinfo {author} {\bibfnamefont {B.}~\bibnamefont {Kannan}}, \bibinfo {author} {\bibfnamefont {A.}~\bibnamefont {Almanakly}}, \bibinfo {author} {\bibfnamefont {Y.}~\bibnamefont {Sung}}, \bibinfo {author} {\bibfnamefont {A.}~\bibnamefont {Di~Paolo}}, \bibinfo {author} {\bibfnamefont {D.~A.}\ \bibnamefont {Rower}}, \bibinfo {author} {\bibfnamefont {J.}~\bibnamefont {Braum{\"u}ller}}, \bibinfo {author} {\bibfnamefont {A.}~\bibnamefont {Melville}}, \bibinfo {author} {\bibfnamefont {B.~M.}\ \bibnamefont {Niedzielski}}, \bibinfo {author} {\bibfnamefont {A.}~\bibnamefont {Karamlou}}, \bibinfo {author} {\bibfnamefont {K.}~\bibnamefont {Serniak}}, \emph {et~al.},\ }\bibfield  {title} {\bibinfo {title} {On-demand directional microwave photon emission using waveguide quantum electrodynamics},\ }\href {https://doi.org/10.1038/s41567-022-01869-5} {\bibfield  {journal} {\bibinfo  {journal} {Nature Physics}\ }\textbf {\bibinfo {volume} {19}},\ \bibinfo {pages} {394} (\bibinfo {year}
  {2023})}\BibitemShut {NoStop}%
\bibitem [{\citenamefont {Lin}\ \emph {et~al.}(2022)\citenamefont {Lin}, \citenamefont {Lu}, \citenamefont {Wen}, \citenamefont {Cheng}, \citenamefont {Lee}, \citenamefont {Lin}, \citenamefont {Chiang}, \citenamefont {Hsieh}, \citenamefont {Chen}, \citenamefont {Chien} \emph {et~al.}}]{lin2022deterministic}%
  \BibitemOpen
  \bibfield  {author} {\bibinfo {author} {\bibfnamefont {W.-J.}\ \bibnamefont {Lin}}, \bibinfo {author} {\bibfnamefont {Y.}~\bibnamefont {Lu}}, \bibinfo {author} {\bibfnamefont {P.~Y.}\ \bibnamefont {Wen}}, \bibinfo {author} {\bibfnamefont {Y.-T.}\ \bibnamefont {Cheng}}, \bibinfo {author} {\bibfnamefont {C.-P.}\ \bibnamefont {Lee}}, \bibinfo {author} {\bibfnamefont {K.~T.}\ \bibnamefont {Lin}}, \bibinfo {author} {\bibfnamefont {K.~H.}\ \bibnamefont {Chiang}}, \bibinfo {author} {\bibfnamefont {M.~C.}\ \bibnamefont {Hsieh}}, \bibinfo {author} {\bibfnamefont {C.-Y.}\ \bibnamefont {Chen}}, \bibinfo {author} {\bibfnamefont {C.-H.}\ \bibnamefont {Chien}}, \emph {et~al.},\ }\bibfield  {title} {\bibinfo {title} {Deterministic loading of microwaves onto an artificial atom using a time-reversed waveform},\ }\href {https://doi.org/10.1021/acs.nanolett.2c02578} {\bibfield  {journal} {\bibinfo  {journal} {Nano Letters}\ }\textbf {\bibinfo {volume} {22}},\ \bibinfo {pages} {8137} (\bibinfo {year} {2022})}\BibitemShut
  {NoStop}%
\bibitem [{\citenamefont {Cheng}\ \emph {et~al.}(2024{\natexlab{a}})\citenamefont {Cheng}, \citenamefont {Chien}, \citenamefont {Hsieh}, \citenamefont {Huang}, \citenamefont {Wen}, \citenamefont {Lin}, \citenamefont {Lu}, \citenamefont {Aziz}, \citenamefont {Lee}, \citenamefont {Lin} \emph {et~al.}}]{cheng2024tuning}%
  \BibitemOpen
  \bibfield  {author} {\bibinfo {author} {\bibfnamefont {Y.-T.}\ \bibnamefont {Cheng}}, \bibinfo {author} {\bibfnamefont {C.-H.}\ \bibnamefont {Chien}}, \bibinfo {author} {\bibfnamefont {K.-M.}\ \bibnamefont {Hsieh}}, \bibinfo {author} {\bibfnamefont {Y.-H.}\ \bibnamefont {Huang}}, \bibinfo {author} {\bibfnamefont {P.}~\bibnamefont {Wen}}, \bibinfo {author} {\bibfnamefont {W.-J.}\ \bibnamefont {Lin}}, \bibinfo {author} {\bibfnamefont {Y.}~\bibnamefont {Lu}}, \bibinfo {author} {\bibfnamefont {F.}~\bibnamefont {Aziz}}, \bibinfo {author} {\bibfnamefont {C.-P.}\ \bibnamefont {Lee}}, \bibinfo {author} {\bibfnamefont {K.-T.}\ \bibnamefont {Lin}}, \emph {et~al.},\ }\bibfield  {title} {\bibinfo {title} {Tuning atom-field interaction via phase shaping},\ }\href {https://doi.org/10.1103/PhysRevA.109.023705} {\bibfield  {journal} {\bibinfo  {journal} {Physical Review A}\ }\textbf {\bibinfo {volume} {109}},\ \bibinfo {pages} {023705} (\bibinfo {year} {2024}{\natexlab{a}})}\BibitemShut {NoStop}%
\bibitem [{\citenamefont {Dorner}\ and\ \citenamefont {Zoller}(2002)}]{Dorner2002}%
  \BibitemOpen
  \bibfield  {author} {\bibinfo {author} {\bibfnamefont {U.}~\bibnamefont {Dorner}}\ and\ \bibinfo {author} {\bibfnamefont {P.}~\bibnamefont {Zoller}},\ }\bibfield  {title} {\bibinfo {title} {Laser-driven atoms in half-cavities},\ }\href {https://doi.org/10.1103/PhysRevA.66.023816} {\bibfield  {journal} {\bibinfo  {journal} {Physical Review A}\ }\textbf {\bibinfo {volume} {66}},\ \bibinfo {pages} {023816} (\bibinfo {year} {2002})}\BibitemShut {NoStop}%
\bibitem [{\citenamefont {Pichler}\ and\ \citenamefont {Zoller}(2016)}]{Pichler2016}%
  \BibitemOpen
  \bibfield  {author} {\bibinfo {author} {\bibfnamefont {H.}~\bibnamefont {Pichler}}\ and\ \bibinfo {author} {\bibfnamefont {P.}~\bibnamefont {Zoller}},\ }\bibfield  {title} {\bibinfo {title} {Photonic {Circuits} with {Time} {Delays} and {Quantum} {Feedback}},\ }\href {https://doi.org/10.1103/PhysRevLett.116.093601} {\bibfield  {journal} {\bibinfo  {journal} {Physical Review Letters}\ }\textbf {\bibinfo {volume} {116}},\ \bibinfo {pages} {093601} (\bibinfo {year} {2016})}\BibitemShut {NoStop}%
\bibitem [{\citenamefont {Zhang}\ \emph {et~al.}(2017)\citenamefont {Zhang}, \citenamefont {Liu}, \citenamefont {Wu}, \citenamefont {Jacobs},\ and\ \citenamefont {Nori}}]{zhang2017quantum}%
  \BibitemOpen
  \bibfield  {author} {\bibinfo {author} {\bibfnamefont {J.}~\bibnamefont {Zhang}}, \bibinfo {author} {\bibfnamefont {Y.-x.}\ \bibnamefont {Liu}}, \bibinfo {author} {\bibfnamefont {R.-B.}\ \bibnamefont {Wu}}, \bibinfo {author} {\bibfnamefont {K.}~\bibnamefont {Jacobs}},\ and\ \bibinfo {author} {\bibfnamefont {F.}~\bibnamefont {Nori}},\ }\bibfield  {title} {\bibinfo {title} {Quantum feedback: {Theory}, experiments, and applications},\ }\href {https://doi.org/10.1016/j.physrep.2017.02.003} {\bibfield  {journal} {\bibinfo  {journal} {Physics Reports}\ }\textbf {\bibinfo {volume} {679}},\ \bibinfo {pages} {1} (\bibinfo {year} {2017})}\BibitemShut {NoStop}%
\bibitem [{\citenamefont {Lu}\ \emph {et~al.}(2017)\citenamefont {Lu}, \citenamefont {Naumann}, \citenamefont {Cerrillo}, \citenamefont {Zhao}, \citenamefont {Knorr},\ and\ \citenamefont {Carmele}}]{Lu2017}%
  \BibitemOpen
  \bibfield  {author} {\bibinfo {author} {\bibfnamefont {Y.}~\bibnamefont {Lu}}, \bibinfo {author} {\bibfnamefont {N.~L.}\ \bibnamefont {Naumann}}, \bibinfo {author} {\bibfnamefont {J.}~\bibnamefont {Cerrillo}}, \bibinfo {author} {\bibfnamefont {Q.}~\bibnamefont {Zhao}}, \bibinfo {author} {\bibfnamefont {A.}~\bibnamefont {Knorr}},\ and\ \bibinfo {author} {\bibfnamefont {A.}~\bibnamefont {Carmele}},\ }\bibfield  {title} {\bibinfo {title} {Intensified antibunching via feedback-induced quantum interference},\ }\href {https://doi.org/10.1103/PhysRevA.95.063840} {\bibfield  {journal} {\bibinfo  {journal} {Physical Review A}\ }\textbf {\bibinfo {volume} {95}},\ \bibinfo {pages} {063840} (\bibinfo {year} {2017})}\BibitemShut {NoStop}%
\bibitem [{\citenamefont {Crowder}\ \emph {et~al.}(2022)\citenamefont {Crowder}, \citenamefont {Ramunno},\ and\ \citenamefont {Hughes}}]{Crowder2022}%
  \BibitemOpen
  \bibfield  {author} {\bibinfo {author} {\bibfnamefont {G.}~\bibnamefont {Crowder}}, \bibinfo {author} {\bibfnamefont {L.}~\bibnamefont {Ramunno}},\ and\ \bibinfo {author} {\bibfnamefont {S.}~\bibnamefont {Hughes}},\ }\bibfield  {title} {\bibinfo {title} {Quantum trajectory theory and simulations of nonlinear spectra and multiphoton effects in waveguide-{QED} systems with a time-delayed coherent feedback},\ }\href {https://doi.org/10.1103/PhysRevA.106.013714} {\bibfield  {journal} {\bibinfo  {journal} {Physical Review A}\ }\textbf {\bibinfo {volume} {106}},\ \bibinfo {pages} {013714} (\bibinfo {year} {2022})}\BibitemShut {NoStop}%
\bibitem [{\citenamefont {Muñoz-Arias}\ \emph {et~al.}(2020)\citenamefont {Muñoz-Arias}, \citenamefont {Poggi}, \citenamefont {Jessen},\ and\ \citenamefont {Deutsch}}]{Munoz-Arias2020}%
  \BibitemOpen
  \bibfield  {author} {\bibinfo {author} {\bibfnamefont {M.~H.}\ \bibnamefont {Muñoz-Arias}}, \bibinfo {author} {\bibfnamefont {P.~M.}\ \bibnamefont {Poggi}}, \bibinfo {author} {\bibfnamefont {P.~S.}\ \bibnamefont {Jessen}},\ and\ \bibinfo {author} {\bibfnamefont {I.~H.}\ \bibnamefont {Deutsch}},\ }\bibfield  {title} {\bibinfo {title} {Simulating {Nonlinear} {Dynamics} of {Collective} {Spins} via {Quantum} {Measurement} and {Feedback}},\ }\href {https://doi.org/10.1103/PhysRevLett.124.110503} {\bibfield  {journal} {\bibinfo  {journal} {Physical Review Letters}\ }\textbf {\bibinfo {volume} {124}},\ \bibinfo {pages} {110503} (\bibinfo {year} {2020})}\BibitemShut {NoStop}%
\bibitem [{\citenamefont {Grigoletto}\ and\ \citenamefont {Ticozzi}(2021)}]{Grigoletto2021}%
  \BibitemOpen
  \bibfield  {author} {\bibinfo {author} {\bibfnamefont {T.}~\bibnamefont {Grigoletto}}\ and\ \bibinfo {author} {\bibfnamefont {F.}~\bibnamefont {Ticozzi}},\ }\bibfield  {title} {\bibinfo {title} {Stabilization {Via} {Feedback} {Switching} for {Quantum} {Stochastic} {Dynamics}},\ }\href {https://doi.org/10.1109/LCSYS.2021.3065603} {\bibfield  {journal} {\bibinfo  {journal} {IEEE Control Systems Letters}\ }\textbf {\bibinfo {volume} {6}},\ \bibinfo {pages} {235} (\bibinfo {year} {2021})}\BibitemShut {NoStop}%
\bibitem [{\citenamefont {Borah}\ \emph {et~al.}(2021)\citenamefont {Borah}, \citenamefont {Sarma}, \citenamefont {Kewming}, \citenamefont {Milburn},\ and\ \citenamefont {Twamley}}]{Borah2021}%
  \BibitemOpen
  \bibfield  {author} {\bibinfo {author} {\bibfnamefont {S.}~\bibnamefont {Borah}}, \bibinfo {author} {\bibfnamefont {B.}~\bibnamefont {Sarma}}, \bibinfo {author} {\bibfnamefont {M.}~\bibnamefont {Kewming}}, \bibinfo {author} {\bibfnamefont {G.~J.}\ \bibnamefont {Milburn}},\ and\ \bibinfo {author} {\bibfnamefont {J.}~\bibnamefont {Twamley}},\ }\bibfield  {title} {\bibinfo {title} {Measurement-{Based} {Feedback} {Quantum} {Control} with {Deep} {Reinforcement} {Learning} for a {Double}-{Well} {Nonlinear} {Potential}},\ }\href {https://doi.org/10.1103/PhysRevLett.127.190403} {\bibfield  {journal} {\bibinfo  {journal} {Physical Review Letters}\ }\textbf {\bibinfo {volume} {127}},\ \bibinfo {pages} {190403} (\bibinfo {year} {2021})}\BibitemShut {NoStop}%
\bibitem [{\citenamefont {Zhang}\ \emph {et~al.}(2012)\citenamefont {Zhang}, \citenamefont {Lo}, \citenamefont {Xiong}, \citenamefont {Tu},\ and\ \citenamefont {Nori}}]{Zhang2012}%
  \BibitemOpen
  \bibfield  {author} {\bibinfo {author} {\bibfnamefont {W.-M.}\ \bibnamefont {Zhang}}, \bibinfo {author} {\bibfnamefont {P.-Y.}\ \bibnamefont {Lo}}, \bibinfo {author} {\bibfnamefont {H.-N.}\ \bibnamefont {Xiong}}, \bibinfo {author} {\bibfnamefont {M.~W.-Y.}\ \bibnamefont {Tu}},\ and\ \bibinfo {author} {\bibfnamefont {F.}~\bibnamefont {Nori}},\ }\bibfield  {title} {\bibinfo {title} {General {Non}-{Markovian} {Dynamics} of {Open} {Quantum} {Systems}},\ }\href {https://doi.org/10.1103/PhysRevLett.109.170402} {\bibfield  {journal} {\bibinfo  {journal} {Physical Review Letters}\ }\textbf {\bibinfo {volume} {109}},\ \bibinfo {pages} {170402} (\bibinfo {year} {2012})}\BibitemShut {NoStop}%
\bibitem [{\citenamefont {Tufarelli}\ \emph {et~al.}(2013)\citenamefont {Tufarelli}, \citenamefont {Ciccarello},\ and\ \citenamefont {Kim}}]{Tufarelli2013}%
  \BibitemOpen
  \bibfield  {author} {\bibinfo {author} {\bibfnamefont {T.}~\bibnamefont {Tufarelli}}, \bibinfo {author} {\bibfnamefont {F.}~\bibnamefont {Ciccarello}},\ and\ \bibinfo {author} {\bibfnamefont {M.~S.}\ \bibnamefont {Kim}},\ }\bibfield  {title} {\bibinfo {title} {Dynamics of spontaneous emission in a single-end photonic waveguide},\ }\href {https://doi.org/10.1103/PhysRevA.87.013820} {\bibfield  {journal} {\bibinfo  {journal} {Physical Review A}\ }\textbf {\bibinfo {volume} {87}},\ \bibinfo {pages} {013820} (\bibinfo {year} {2013})}\BibitemShut {NoStop}%
\bibitem [{\citenamefont {Carmele}\ \emph {et~al.}(2013)\citenamefont {Carmele}, \citenamefont {Kabuss}, \citenamefont {Schulze}, \citenamefont {Reitzenstein},\ and\ \citenamefont {Knorr}}]{Carmele2013}%
  \BibitemOpen
  \bibfield  {author} {\bibinfo {author} {\bibfnamefont {A.}~\bibnamefont {Carmele}}, \bibinfo {author} {\bibfnamefont {J.}~\bibnamefont {Kabuss}}, \bibinfo {author} {\bibfnamefont {F.}~\bibnamefont {Schulze}}, \bibinfo {author} {\bibfnamefont {S.}~\bibnamefont {Reitzenstein}},\ and\ \bibinfo {author} {\bibfnamefont {A.}~\bibnamefont {Knorr}},\ }\bibfield  {title} {\bibinfo {title} {Single {Photon} {Delayed} {Feedback}: {A} {Way} to {Stabilize} {Intrinsic} {Quantum} {Cavity} {Electrodynamics}},\ }\href {https://doi.org/10.1103/PhysRevLett.110.013601} {\bibfield  {journal} {\bibinfo  {journal} {Physical Review Letters}\ }\textbf {\bibinfo {volume} {110}},\ \bibinfo {pages} {013601} (\bibinfo {year} {2013})}\BibitemShut {NoStop}%
\bibitem [{\citenamefont {Grimsmo}(2015)}]{Grimsmo2015}%
  \BibitemOpen
  \bibfield  {author} {\bibinfo {author} {\bibfnamefont {A.~L.}\ \bibnamefont {Grimsmo}},\ }\bibfield  {title} {\bibinfo {title} {Time-{Delayed} {Quantum} {Feedback} {Control}},\ }\href {https://doi.org/10.1103/PhysRevLett.115.060402} {\bibfield  {journal} {\bibinfo  {journal} {Physical Review Letters}\ }\textbf {\bibinfo {volume} {115}},\ \bibinfo {pages} {060402} (\bibinfo {year} {2015})}\BibitemShut {NoStop}%
\bibitem [{\citenamefont {N\'{e}met}\ and\ \citenamefont {Parkins}(2016)}]{Nemet2016}%
  \BibitemOpen
  \bibfield  {author} {\bibinfo {author} {\bibfnamefont {N.}~\bibnamefont {N\'{e}met}}\ and\ \bibinfo {author} {\bibfnamefont {S.}~\bibnamefont {Parkins}},\ }\bibfield  {title} {\bibinfo {title} {Enhanced optical squeezing from a degenerate parametric amplifier via time-delayed coherent feedback},\ }\href {https://doi.org/10.1103/PhysRevA.94.023809} {\bibfield  {journal} {\bibinfo  {journal} {Physical Review A}\ }\textbf {\bibinfo {volume} {94}},\ \bibinfo {pages} {023809} (\bibinfo {year} {2016})}\BibitemShut {NoStop}%
\bibitem [{\citenamefont {Chen}\ \emph {et~al.}(2016)\citenamefont {Chen}, \citenamefont {Lambert}, \citenamefont {Li}, \citenamefont {Miranowicz}, \citenamefont {Chen},\ and\ \citenamefont {Nori}}]{Chen2016}%
  \BibitemOpen
  \bibfield  {author} {\bibinfo {author} {\bibfnamefont {S.-L.}\ \bibnamefont {Chen}}, \bibinfo {author} {\bibfnamefont {N.}~\bibnamefont {Lambert}}, \bibinfo {author} {\bibfnamefont {C.-M.}\ \bibnamefont {Li}}, \bibinfo {author} {\bibfnamefont {A.}~\bibnamefont {Miranowicz}}, \bibinfo {author} {\bibfnamefont {Y.-N.}\ \bibnamefont {Chen}},\ and\ \bibinfo {author} {\bibfnamefont {F.}~\bibnamefont {Nori}},\ }\bibfield  {title} {\bibinfo {title} {Quantifying {Non}-{Markovianity} with {Temporal} {Steering}},\ }\href {https://doi.org/10.1103/PhysRevLett.116.020503} {\bibfield  {journal} {\bibinfo  {journal} {Physical Review Letters}\ }\textbf {\bibinfo {volume} {116}},\ \bibinfo {pages} {020503} (\bibinfo {year} {2016})}\BibitemShut {NoStop}%
\bibitem [{\citenamefont {Pichler}\ \emph {et~al.}(2017)\citenamefont {Pichler}, \citenamefont {Choi}, \citenamefont {Zoller},\ and\ \citenamefont {Lukin}}]{Pichler2017}%
  \BibitemOpen
  \bibfield  {author} {\bibinfo {author} {\bibfnamefont {H.}~\bibnamefont {Pichler}}, \bibinfo {author} {\bibfnamefont {S.}~\bibnamefont {Choi}}, \bibinfo {author} {\bibfnamefont {P.}~\bibnamefont {Zoller}},\ and\ \bibinfo {author} {\bibfnamefont {M.~D.}\ \bibnamefont {Lukin}},\ }\bibfield  {title} {\bibinfo {title} {Universal photonic quantum computation via time-delayed feedback},\ }\href {https://doi.org/10.1073/pnas.1711003114} {\bibfield  {journal} {\bibinfo  {journal} {Proceedings of the National Academy of Sciences}\ }\textbf {\bibinfo {volume} {114}},\ \bibinfo {pages} {11362} (\bibinfo {year} {2017})}\BibitemShut {NoStop}%
\bibitem [{\citenamefont {Whalen}\ \emph {et~al.}(2017)\citenamefont {Whalen}, \citenamefont {Grimsmo},\ and\ \citenamefont {Carmichael}}]{Whalen2017}%
  \BibitemOpen
  \bibfield  {author} {\bibinfo {author} {\bibfnamefont {S.~J.}\ \bibnamefont {Whalen}}, \bibinfo {author} {\bibfnamefont {A.~L.}\ \bibnamefont {Grimsmo}},\ and\ \bibinfo {author} {\bibfnamefont {H.~J.}\ \bibnamefont {Carmichael}},\ }\bibfield  {title} {\bibinfo {title} {Open quantum systems with delayed coherent feedback},\ }\href {https://doi.org/10.1088/2058-9565/aa8331} {\bibfield  {journal} {\bibinfo  {journal} {Quantum Science and Technology}\ }\textbf {\bibinfo {volume} {2}},\ \bibinfo {pages} {044008} (\bibinfo {year} {2017})}\BibitemShut {NoStop}%
\bibitem [{\citenamefont {Andersson}\ \emph {et~al.}(2019)\citenamefont {Andersson}, \citenamefont {Suri}, \citenamefont {Guo}, \citenamefont {Aref},\ and\ \citenamefont {Delsing}}]{andersson2019non}%
  \BibitemOpen
  \bibfield  {author} {\bibinfo {author} {\bibfnamefont {G.}~\bibnamefont {Andersson}}, \bibinfo {author} {\bibfnamefont {B.}~\bibnamefont {Suri}}, \bibinfo {author} {\bibfnamefont {L.}~\bibnamefont {Guo}}, \bibinfo {author} {\bibfnamefont {T.}~\bibnamefont {Aref}},\ and\ \bibinfo {author} {\bibfnamefont {P.}~\bibnamefont {Delsing}},\ }\bibfield  {title} {\bibinfo {title} {Non-exponential decay of a giant artificial atom},\ }\href {https://doi.org/10.1038/s41567-019-0605-6} {\bibfield  {journal} {\bibinfo  {journal} {Nature Physics}\ }\textbf {\bibinfo {volume} {15}},\ \bibinfo {pages} {1123} (\bibinfo {year} {2019})}\BibitemShut {NoStop}%
\bibitem [{\citenamefont {Calaj\'{o}}\ \emph {et~al.}(2019)\citenamefont {Calaj\'{o}}, \citenamefont {Fang}, \citenamefont {Baranger},\ and\ \citenamefont {Ciccarello}}]{Calajo2019}%
  \BibitemOpen
  \bibfield  {author} {\bibinfo {author} {\bibfnamefont {G.}~\bibnamefont {Calaj\'{o}}}, \bibinfo {author} {\bibfnamefont {Y.-L.~L.}\ \bibnamefont {Fang}}, \bibinfo {author} {\bibfnamefont {H.~U.}\ \bibnamefont {Baranger}},\ and\ \bibinfo {author} {\bibfnamefont {F.}~\bibnamefont {Ciccarello}},\ }\bibfield  {title} {\bibinfo {title} {Exciting a {Bound} {State} in the {Continuum} through {Multiphoton} {Scattering} {Plus} {Delayed} {Quantum} {Feedback}},\ }\href {https://doi.org/10.1103/PhysRevLett.122.073601} {\bibfield  {journal} {\bibinfo  {journal} {Physical Review Letters}\ }\textbf {\bibinfo {volume} {122}},\ \bibinfo {pages} {073601} (\bibinfo {year} {2019})}\BibitemShut {NoStop}%
\bibitem [{\citenamefont {Crowder}\ \emph {et~al.}(2020)\citenamefont {Crowder}, \citenamefont {Carmichael},\ and\ \citenamefont {Hughes}}]{Crowder2020}%
  \BibitemOpen
  \bibfield  {author} {\bibinfo {author} {\bibfnamefont {G.}~\bibnamefont {Crowder}}, \bibinfo {author} {\bibfnamefont {H.~J.}\ \bibnamefont {Carmichael}},\ and\ \bibinfo {author} {\bibfnamefont {S.}~\bibnamefont {Hughes}},\ }\bibfield  {title} {\bibinfo {title} {Quantum trajectory theory of few-photon cavity-{QED} systems with a time-delayed coherent feedback},\ }\href {https://doi.org/10.1103/PhysRevA.101.023807} {\bibfield  {journal} {\bibinfo  {journal} {Physical Review A}\ }\textbf {\bibinfo {volume} {101}},\ \bibinfo {pages} {023807} (\bibinfo {year} {2020})}\BibitemShut {NoStop}%
\bibitem [{\citenamefont {Lu}\ \emph {et~al.}(2020)\citenamefont {Lu}, \citenamefont {Zhang}, \citenamefont {Liu}, \citenamefont {Nori}, \citenamefont {Fan},\ and\ \citenamefont {Pan}}]{Lu2020}%
  \BibitemOpen
  \bibfield  {author} {\bibinfo {author} {\bibfnamefont {Y.-N.}\ \bibnamefont {Lu}}, \bibinfo {author} {\bibfnamefont {Y.-R.}\ \bibnamefont {Zhang}}, \bibinfo {author} {\bibfnamefont {G.-Q.}\ \bibnamefont {Liu}}, \bibinfo {author} {\bibfnamefont {F.}~\bibnamefont {Nori}}, \bibinfo {author} {\bibfnamefont {H.}~\bibnamefont {Fan}},\ and\ \bibinfo {author} {\bibfnamefont {X.-Y.}\ \bibnamefont {Pan}},\ }\bibfield  {title} {\bibinfo {title} {Observing {Information} {Backflow} from {Controllable} {Non}-{Markovian} {Multichannels} in {Diamond}},\ }\href {https://doi.org/10.1103/PhysRevLett.124.210502} {\bibfield  {journal} {\bibinfo  {journal} {Physical Review Letters}\ }\textbf {\bibinfo {volume} {124}},\ \bibinfo {pages} {210502} (\bibinfo {year} {2020})}\BibitemShut {NoStop}%
\bibitem [{\citenamefont {Arranz~Regidor}\ \emph {et~al.}(2021)\citenamefont {Arranz~Regidor}, \citenamefont {Crowder}, \citenamefont {Carmichael},\ and\ \citenamefont {Hughes}}]{Regidor2021a}%
  \BibitemOpen
  \bibfield  {author} {\bibinfo {author} {\bibfnamefont {S.}~\bibnamefont {Arranz~Regidor}}, \bibinfo {author} {\bibfnamefont {G.}~\bibnamefont {Crowder}}, \bibinfo {author} {\bibfnamefont {H.~J.}\ \bibnamefont {Carmichael}},\ and\ \bibinfo {author} {\bibfnamefont {S.}~\bibnamefont {Hughes}},\ }\bibfield  {title} {\bibinfo {title} {Modeling quantum light-matter interactions in waveguide {QED} with retardation, nonlinear interactions, and a time-delayed feedback: {Matrix} product states versus a space-discretized waveguide model},\ }\href {https://doi.org/10.1103/PhysRevResearch.3.023030} {\bibfield  {journal} {\bibinfo  {journal} {Physical Review Research}\ }\textbf {\bibinfo {volume} {3}},\ \bibinfo {pages} {023030} (\bibinfo {year} {2021})}\BibitemShut {NoStop}%
\bibitem [{\citenamefont {Barkemeyer}\ \emph {et~al.}(2022)\citenamefont {Barkemeyer}, \citenamefont {Knorr},\ and\ \citenamefont {Carmele}}]{Barkemeyer2022}%
  \BibitemOpen
  \bibfield  {author} {\bibinfo {author} {\bibfnamefont {K.}~\bibnamefont {Barkemeyer}}, \bibinfo {author} {\bibfnamefont {A.}~\bibnamefont {Knorr}},\ and\ \bibinfo {author} {\bibfnamefont {A.}~\bibnamefont {Carmele}},\ }\bibfield  {title} {\bibinfo {title} {Heisenberg treatment of multiphoton pulses in waveguide {QED} with time-delayed feedback},\ }\href {https://doi.org/10.1103/PhysRevA.106.023708} {\bibfield  {journal} {\bibinfo  {journal} {Physical Review A}\ }\textbf {\bibinfo {volume} {106}},\ \bibinfo {pages} {023708} (\bibinfo {year} {2022})}\BibitemShut {NoStop}%
\bibitem [{\citenamefont {Crowder}\ \emph {et~al.}(2024)\citenamefont {Crowder}, \citenamefont {Ramunno},\ and\ \citenamefont {Hughes}}]{Crowder2024}%
  \BibitemOpen
  \bibfield  {author} {\bibinfo {author} {\bibfnamefont {G.}~\bibnamefont {Crowder}}, \bibinfo {author} {\bibfnamefont {L.}~\bibnamefont {Ramunno}},\ and\ \bibinfo {author} {\bibfnamefont {S.}~\bibnamefont {Hughes}},\ }\bibfield  {title} {\bibinfo {title} {Improving on-demand single-photon-source coherence and indistinguishability through a time-delayed coherent feedback},\ }\href {https://doi.org/10.1103/PhysRevA.110.L031703} {\bibfield  {journal} {\bibinfo  {journal} {Physical Review A}\ }\textbf {\bibinfo {volume} {110}},\ \bibinfo {pages} {L031703} (\bibinfo {year} {2024})}\BibitemShut {NoStop}%
\bibitem [{\citenamefont {Vodenkova}\ and\ \citenamefont {Pichler}(2024)}]{Vodenkova2024}%
  \BibitemOpen
  \bibfield  {author} {\bibinfo {author} {\bibfnamefont {K.}~\bibnamefont {Vodenkova}}\ and\ \bibinfo {author} {\bibfnamefont {H.}~\bibnamefont {Pichler}},\ }\bibfield  {title} {\bibinfo {title} {Continuous coherent quantum feedback with time delays: Tensor network solution},\ }\href {https://doi.org/10.1103/PhysRevX.14.031043} {\bibfield  {journal} {\bibinfo  {journal} {Physical Review X}\ }\textbf {\bibinfo {volume} {14}},\ \bibinfo {pages} {031043} (\bibinfo {year} {2024})}\BibitemShut {NoStop}%
\bibitem [{\citenamefont {Ferreira}\ \emph {et~al.}(2021)\citenamefont {Ferreira}, \citenamefont {Banker}, \citenamefont {Sipahigil}, \citenamefont {Matheny}, \citenamefont {Keller}, \citenamefont {Kim}, \citenamefont {Mirhosseini},\ and\ \citenamefont {Painter}}]{ferreira2021collapse}%
  \BibitemOpen
  \bibfield  {author} {\bibinfo {author} {\bibfnamefont {V.~S.}\ \bibnamefont {Ferreira}}, \bibinfo {author} {\bibfnamefont {J.}~\bibnamefont {Banker}}, \bibinfo {author} {\bibfnamefont {A.}~\bibnamefont {Sipahigil}}, \bibinfo {author} {\bibfnamefont {M.~H.}\ \bibnamefont {Matheny}}, \bibinfo {author} {\bibfnamefont {A.~J.}\ \bibnamefont {Keller}}, \bibinfo {author} {\bibfnamefont {E.}~\bibnamefont {Kim}}, \bibinfo {author} {\bibfnamefont {M.}~\bibnamefont {Mirhosseini}},\ and\ \bibinfo {author} {\bibfnamefont {O.}~\bibnamefont {Painter}},\ }\bibfield  {title} {\bibinfo {title} {Collapse and revival of an artificial atom coupled to a structured photonic reservoir},\ }\href {https://journals.aps.org/prx/abstract/10.1103/PhysRevX.11.041043} {\bibfield  {journal} {\bibinfo  {journal} {Physical Review X}\ }\textbf {\bibinfo {volume} {11}},\ \bibinfo {pages} {041043} (\bibinfo {year} {2021})}\BibitemShut {NoStop}%
\bibitem [{\citenamefont {Ferreira}\ \emph {et~al.}(2024)\citenamefont {Ferreira}, \citenamefont {Kim}, \citenamefont {Butler}, \citenamefont {Pichler},\ and\ \citenamefont {Painter}}]{Ferreira2024}%
  \BibitemOpen
  \bibfield  {author} {\bibinfo {author} {\bibfnamefont {V.~S.}\ \bibnamefont {Ferreira}}, \bibinfo {author} {\bibfnamefont {G.}~\bibnamefont {Kim}}, \bibinfo {author} {\bibfnamefont {A.}~\bibnamefont {Butler}}, \bibinfo {author} {\bibfnamefont {H.}~\bibnamefont {Pichler}},\ and\ \bibinfo {author} {\bibfnamefont {O.}~\bibnamefont {Painter}},\ }\bibfield  {title} {\bibinfo {title} {Deterministic generation of multidimensional photonic cluster states with a single quantum emitter},\ }\href {https://doi.org/10.1038/s41567-024-02408-0} {\bibfield  {journal} {\bibinfo  {journal} {Nature Physics}\ }\textbf {\bibinfo {volume} {20}},\ \bibinfo {pages} {865} (\bibinfo {year} {2024})}\BibitemShut {NoStop}%
\bibitem [{\citenamefont {Bienfait}\ \emph {et~al.}(2019)\citenamefont {Bienfait}, \citenamefont {Satzinger}, \citenamefont {Zhong}, \citenamefont {Chang}, \citenamefont {Chou}, \citenamefont {Conner}, \citenamefont {Dumur}, \citenamefont {Grebel}, \citenamefont {Peairs}, \citenamefont {Povey} \emph {et~al.}}]{bienfait2019phonon}%
  \BibitemOpen
  \bibfield  {author} {\bibinfo {author} {\bibfnamefont {A.}~\bibnamefont {Bienfait}}, \bibinfo {author} {\bibfnamefont {K.~J.}\ \bibnamefont {Satzinger}}, \bibinfo {author} {\bibfnamefont {Y.}~\bibnamefont {Zhong}}, \bibinfo {author} {\bibfnamefont {H.-S.}\ \bibnamefont {Chang}}, \bibinfo {author} {\bibfnamefont {M.-H.}\ \bibnamefont {Chou}}, \bibinfo {author} {\bibfnamefont {C.~R.}\ \bibnamefont {Conner}}, \bibinfo {author} {\bibfnamefont {{\'E}.}~\bibnamefont {Dumur}}, \bibinfo {author} {\bibfnamefont {J.}~\bibnamefont {Grebel}}, \bibinfo {author} {\bibfnamefont {G.~A.}\ \bibnamefont {Peairs}}, \bibinfo {author} {\bibfnamefont {R.~G.}\ \bibnamefont {Povey}}, \emph {et~al.},\ }\bibfield  {title} {\bibinfo {title} {Phonon-mediated quantum state transfer and remote qubit entanglement},\ }\href {https://www.science.org/doi/10.1126/science.aaw8415} {\bibfield  {journal} {\bibinfo  {journal} {Science}\ }\textbf {\bibinfo {volume} {364}},\ \bibinfo {pages} {368} (\bibinfo {year} {2019})}\BibitemShut {NoStop}%
\bibitem [{\citenamefont {Zhong}\ \emph {et~al.}(2019)\citenamefont {Zhong}, \citenamefont {Chang}, \citenamefont {Satzinger}, \citenamefont {Chou}, \citenamefont {Bienfait}, \citenamefont {Conner}, \citenamefont {Dumur}, \citenamefont {Grebel}, \citenamefont {Peairs}, \citenamefont {Povey} \emph {et~al.}}]{zhong2019violating}%
  \BibitemOpen
  \bibfield  {author} {\bibinfo {author} {\bibfnamefont {Y.}~\bibnamefont {Zhong}}, \bibinfo {author} {\bibfnamefont {H.-S.}\ \bibnamefont {Chang}}, \bibinfo {author} {\bibfnamefont {K.}~\bibnamefont {Satzinger}}, \bibinfo {author} {\bibfnamefont {M.-H.}\ \bibnamefont {Chou}}, \bibinfo {author} {\bibfnamefont {A.}~\bibnamefont {Bienfait}}, \bibinfo {author} {\bibfnamefont {C.}~\bibnamefont {Conner}}, \bibinfo {author} {\bibfnamefont {{\'E}.}~\bibnamefont {Dumur}}, \bibinfo {author} {\bibfnamefont {J.}~\bibnamefont {Grebel}}, \bibinfo {author} {\bibfnamefont {G.}~\bibnamefont {Peairs}}, \bibinfo {author} {\bibfnamefont {R.}~\bibnamefont {Povey}}, \emph {et~al.},\ }\bibfield  {title} {\bibinfo {title} {Violating bell’s inequality with remotely connected superconducting qubits},\ }\href {https://www.nature.com/articles/s41567-019-0507-7} {\bibfield  {journal} {\bibinfo  {journal} {Nature Physics}\ }\textbf {\bibinfo {volume} {15}},\ \bibinfo {pages} {741} (\bibinfo {year} {2019})}\BibitemShut {NoStop}%
\bibitem [{\citenamefont {Odeh}\ \emph {et~al.}(2025)\citenamefont {Odeh}, \citenamefont {Godeneli}, \citenamefont {Li}, \citenamefont {Tangirala}, \citenamefont {Zhou}, \citenamefont {Zhang}, \citenamefont {Zhang},\ and\ \citenamefont {Sipahigil}}]{odeh2025non}%
  \BibitemOpen
  \bibfield  {author} {\bibinfo {author} {\bibfnamefont {M.}~\bibnamefont {Odeh}}, \bibinfo {author} {\bibfnamefont {K.}~\bibnamefont {Godeneli}}, \bibinfo {author} {\bibfnamefont {E.}~\bibnamefont {Li}}, \bibinfo {author} {\bibfnamefont {R.}~\bibnamefont {Tangirala}}, \bibinfo {author} {\bibfnamefont {H.}~\bibnamefont {Zhou}}, \bibinfo {author} {\bibfnamefont {X.}~\bibnamefont {Zhang}}, \bibinfo {author} {\bibfnamefont {Z.-H.}\ \bibnamefont {Zhang}},\ and\ \bibinfo {author} {\bibfnamefont {A.}~\bibnamefont {Sipahigil}},\ }\bibfield  {title} {\bibinfo {title} {Non-markovian dynamics of a superconducting qubit in a phononic bandgap},\ }\href {https://www.nature.com/articles/s41567-024-02740-5} {\bibfield  {journal} {\bibinfo  {journal} {Nature Physics}\ }\textbf {\bibinfo {volume} {21}},\ \bibinfo {pages} {406} (\bibinfo {year} {2025})}\BibitemShut {NoStop}%
\bibitem [{\citenamefont {Gardiner}\ and\ \citenamefont {Zoller}(2000)}]{QuantumNoise}%
  \BibitemOpen
  \bibfield  {author} {\bibinfo {author} {\bibfnamefont {C.}~\bibnamefont {Gardiner}}\ and\ \bibinfo {author} {\bibfnamefont {P.}~\bibnamefont {Zoller}},\ }\href@noop {} {\emph {\bibinfo {title} {Quantum Noise: A Handbook of Markovian and Non-Markovian Quantum Stochastic Methods with Applications to Quantum Optics}}}\ (\bibinfo  {publisher} {Springer-Verlag},\ \bibinfo {address} {Berlin},\ \bibinfo {year} {2000})\BibitemShut {NoStop}%
\bibitem [{\citenamefont {Stenius}\ and\ \citenamefont {Imamoglu}(1996)}]{Stenius1996}%
  \BibitemOpen
  \bibfield  {author} {\bibinfo {author} {\bibfnamefont {P.}~\bibnamefont {Stenius}}\ and\ \bibinfo {author} {\bibfnamefont {A.}~\bibnamefont {Imamoglu}},\ }\bibfield  {title} {\bibinfo {title} {Stochastic wavefunction methods beyond the {Born}-{Markov} and rotating-wave approximations},\ }\href {https://doi.org/10.1088/1355-5111/8/1/021} {\bibfield  {journal} {\bibinfo  {journal} {Quantum and Semiclassical Optics: Journal of the European Optical Society Part B}\ }\textbf {\bibinfo {volume} {8}},\ \bibinfo {pages} {283} (\bibinfo {year} {1996})}\BibitemShut {NoStop}%
\bibitem [{\citenamefont {Whalen}(2019)}]{Whalen2019}%
  \BibitemOpen
  \bibfield  {author} {\bibinfo {author} {\bibfnamefont {S.~J.}\ \bibnamefont {Whalen}},\ }\bibfield  {title} {\bibinfo {title} {Collision model for non-{Markovian} quantum trajectories},\ }\href {https://doi.org/10.1103/PhysRevA.100.052113} {\bibfield  {journal} {\bibinfo  {journal} {Physical Review A}\ }\textbf {\bibinfo {volume} {100}},\ \bibinfo {pages} {052113} (\bibinfo {year} {2019})}\BibitemShut {NoStop}%
\bibitem [{Sup()}]{SuppMaterial}%
  \BibitemOpen
  \href@noop {} {}\bibinfo {note} {Supplementary material}\BibitemShut {NoStop}%
\bibitem [{\citenamefont {Koch}\ \emph {et~al.}(2007)\citenamefont {Koch}, \citenamefont {Yu}, \citenamefont {Gambetta}, \citenamefont {Houck}, \citenamefont {Schuster}, \citenamefont {Majer}, \citenamefont {Blais}, \citenamefont {Devoret}, \citenamefont {Girvin},\ and\ \citenamefont {Schoelkopf}}]{koch2007charge}%
  \BibitemOpen
  \bibfield  {author} {\bibinfo {author} {\bibfnamefont {J.}~\bibnamefont {Koch}}, \bibinfo {author} {\bibfnamefont {T.~M.}\ \bibnamefont {Yu}}, \bibinfo {author} {\bibfnamefont {J.}~\bibnamefont {Gambetta}}, \bibinfo {author} {\bibfnamefont {A.~A.}\ \bibnamefont {Houck}}, \bibinfo {author} {\bibfnamefont {D.~I.}\ \bibnamefont {Schuster}}, \bibinfo {author} {\bibfnamefont {J.}~\bibnamefont {Majer}}, \bibinfo {author} {\bibfnamefont {A.}~\bibnamefont {Blais}}, \bibinfo {author} {\bibfnamefont {M.~H.}\ \bibnamefont {Devoret}}, \bibinfo {author} {\bibfnamefont {S.~M.}\ \bibnamefont {Girvin}},\ and\ \bibinfo {author} {\bibfnamefont {R.~J.}\ \bibnamefont {Schoelkopf}},\ }\bibfield  {title} {\bibinfo {title} {Charge-insensitive qubit design derived from the {C}ooper pair box},\ }\href {https://doi.org/10.1103/PhysRevA.76.042319} {\bibfield  {journal} {\bibinfo  {journal} {Physical Review A}\ }\textbf {\bibinfo {volume} {76}},\ \bibinfo {pages} {042319} (\bibinfo {year} {2007})}\BibitemShut {NoStop}%
\bibitem [{\citenamefont {Aziz}\ \emph {et~al.}(2025)\citenamefont {Aziz}, \citenamefont {Lin}, \citenamefont {Wen}, \citenamefont {Lin}, \citenamefont {Weigand}, \citenamefont {Lee}, \citenamefont {Cheng}, \citenamefont {Lu}, \citenamefont {Chen}, \citenamefont {Chien} \emph {et~al.}}]{aziz2025nearly}%
  \BibitemOpen
  \bibfield  {author} {\bibinfo {author} {\bibfnamefont {F.}~\bibnamefont {Aziz}}, \bibinfo {author} {\bibfnamefont {K.-T.}\ \bibnamefont {Lin}}, \bibinfo {author} {\bibfnamefont {P.-Y.}\ \bibnamefont {Wen}}, \bibinfo {author} {\bibfnamefont {Y.-C.}\ \bibnamefont {Lin}}, \bibinfo {author} {\bibfnamefont {E.}~\bibnamefont {Weigand}}, \bibinfo {author} {\bibfnamefont {C.-P.}\ \bibnamefont {Lee}}, \bibinfo {author} {\bibfnamefont {Y.-T.}\ \bibnamefont {Cheng}}, \bibinfo {author} {\bibfnamefont {Y.}~\bibnamefont {Lu}}, \bibinfo {author} {\bibfnamefont {C.-Y.}\ \bibnamefont {Chen}}, \bibinfo {author} {\bibfnamefont {C.-H.}\ \bibnamefont {Chien}}, \emph {et~al.},\ }\bibfield  {title} {\bibinfo {title} {Nearly quantum-limited microwave amplification via interfering degenerate stimulated emission in a single artificial atom},\ }\href {https://doi.org/10.1038/s41534-025-00993-3} {\bibfield  {journal} {\bibinfo  {journal} {npj Quantum Information}\ }\textbf {\bibinfo {volume} {11}},\ \bibinfo {pages} {1} (\bibinfo
  {year} {2025})}\BibitemShut {NoStop}%
\bibitem [{en1()}]{en1}%
  \BibitemOpen
  \href@noop {} {}\bibinfo {note} {In the very strong coupling, any low-$Q$ stray resonance mode along the waveguide will affect the data. {\color{black}It can be caused by the Fano effect along the transmission line.}}\BibitemShut {Stop}%
\bibitem [{\citenamefont {Koshino}\ and\ \citenamefont {Nakamura}(2012)}]{koshino2012control}%
  \BibitemOpen
  \bibfield  {author} {\bibinfo {author} {\bibfnamefont {K.}~\bibnamefont {Koshino}}\ and\ \bibinfo {author} {\bibfnamefont {Y.}~\bibnamefont {Nakamura}},\ }\bibfield  {title} {\bibinfo {title} {Control of the radiative level shift and linewidth of a superconducting artificial atom through a variable boundary condition},\ }\href {https://doi.org/10.1088/1367-2630/14/4/043005} {\bibfield  {journal} {\bibinfo  {journal} {New Journal of Physics}\ }\textbf {\bibinfo {volume} {14}},\ \bibinfo {pages} {043005} (\bibinfo {year} {2012})}\BibitemShut {NoStop}%
\bibitem [{\citenamefont {Wu}\ \emph {et~al.}(2024)\citenamefont {Wu}, \citenamefont {Cheng}, \citenamefont {Lin}, \citenamefont {Aziz}, \citenamefont {Liu}, \citenamefont {Rangdhol}, \citenamefont {Yeung}, \citenamefont {Yang}, \citenamefont {Shao}, \citenamefont {Wang} \emph {et~al.}}]{wu2024microwave}%
  \BibitemOpen
  \bibfield  {author} {\bibinfo {author} {\bibfnamefont {B.-Y.}\ \bibnamefont {Wu}}, \bibinfo {author} {\bibfnamefont {Y.-T.}\ \bibnamefont {Cheng}}, \bibinfo {author} {\bibfnamefont {K.-T.}\ \bibnamefont {Lin}}, \bibinfo {author} {\bibfnamefont {F.}~\bibnamefont {Aziz}}, \bibinfo {author} {\bibfnamefont {J.-C.}\ \bibnamefont {Liu}}, \bibinfo {author} {\bibfnamefont {K.-V.}\ \bibnamefont {Rangdhol}}, \bibinfo {author} {\bibfnamefont {Y.-Y.}\ \bibnamefont {Yeung}}, \bibinfo {author} {\bibfnamefont {S.}~\bibnamefont {Yang}}, \bibinfo {author} {\bibfnamefont {Q.}~\bibnamefont {Shao}}, \bibinfo {author} {\bibfnamefont {X.}~\bibnamefont {Wang}}, \emph {et~al.},\ }\bibfield  {title} {\bibinfo {title} {Microwave interference from a spin ensemble and its mirror image in waveguide magnonics},\ }\href {https://doi.org/10.48550/arXiv.2409.17867} {\bibfield  {journal} {\bibinfo  {journal} {arXiv preprint arXiv:2409.17867}\ } (\bibinfo {year} {2024})}\BibitemShut {NoStop}%
\bibitem [{\citenamefont {Cheng}\ \emph {et~al.}(2024{\natexlab{b}})\citenamefont {Cheng}, \citenamefont {Hsieh}, \citenamefont {Wu}, \citenamefont {Niu}, \citenamefont {Aziz}, \citenamefont {Huang}, \citenamefont {Wen}, \citenamefont {Lin}, \citenamefont {Lin}, \citenamefont {Chen} \emph {et~al.}}]{cheng2024group}%
  \BibitemOpen
  \bibfield  {author} {\bibinfo {author} {\bibfnamefont {Y.-T.}\ \bibnamefont {Cheng}}, \bibinfo {author} {\bibfnamefont {K.-M.}\ \bibnamefont {Hsieh}}, \bibinfo {author} {\bibfnamefont {B.-Y.}\ \bibnamefont {Wu}}, \bibinfo {author} {\bibfnamefont {Z.}~\bibnamefont {Niu}}, \bibinfo {author} {\bibfnamefont {F.}~\bibnamefont {Aziz}}, \bibinfo {author} {\bibfnamefont {Y.-H.}\ \bibnamefont {Huang}}, \bibinfo {author} {\bibfnamefont {P.}~\bibnamefont {Wen}}, \bibinfo {author} {\bibfnamefont {K.-T.}\ \bibnamefont {Lin}}, \bibinfo {author} {\bibfnamefont {Y.-H.}\ \bibnamefont {Lin}}, \bibinfo {author} {\bibfnamefont {J.}~\bibnamefont {Chen}}, \emph {et~al.},\ }\bibfield  {title} {\bibinfo {title} {Group delay controlled by the decoherence of a single artificial atom},\ }\href {https://doi.org/10.48550/arXiv.2409.07731} {\bibfield  {journal} {\bibinfo  {journal} {arXiv preprint arXiv:2409.07731}\ } (\bibinfo {year} {2024}{\natexlab{b}})}\BibitemShut {NoStop}%
\bibitem [{\citenamefont {Probst}\ \emph {et~al.}(2015)\citenamefont {Probst}, \citenamefont {Song}, \citenamefont {Bushev}, \citenamefont {Ustinov},\ and\ \citenamefont {Weides}}]{probst2015efficient}%
  \BibitemOpen
  \bibfield  {author} {\bibinfo {author} {\bibfnamefont {S.}~\bibnamefont {Probst}}, \bibinfo {author} {\bibfnamefont {F.}~\bibnamefont {Song}}, \bibinfo {author} {\bibfnamefont {P.~A.}\ \bibnamefont {Bushev}}, \bibinfo {author} {\bibfnamefont {A.~V.}\ \bibnamefont {Ustinov}},\ and\ \bibinfo {author} {\bibfnamefont {M.}~\bibnamefont {Weides}},\ }\bibfield  {title} {\bibinfo {title} {Efficient and robust analysis of complex scattering data under noise in microwave resonators},\ }\href {https://doi.org/10.1063/1.4907935} {\bibfield  {journal} {\bibinfo  {journal} {Review of Scientific Instruments}\ }\textbf {\bibinfo {volume} {86}} (\bibinfo {year} {2015})}\BibitemShut {NoStop}%
\bibitem [{\citenamefont {Cheng}\ \emph {et~al.}(2025)\citenamefont {Cheng}, \citenamefont {Hsieh}, \citenamefont {Wu}, \citenamefont {Niu}, \citenamefont {Aziz}, \citenamefont {Huang}, \citenamefont {Wen}, \citenamefont {Lin}, \citenamefont {Lin}, \citenamefont {Chen}, \citenamefont {Kockum}, \citenamefont {Lin}, \citenamefont {Lin}, \citenamefont {Lu},\ and\ \citenamefont {Hoi}}]{Cheng2025}%
  \BibitemOpen
  \bibfield  {author} {\bibinfo {author} {\bibfnamefont {Y.-T.}\ \bibnamefont {Cheng}}, \bibinfo {author} {\bibfnamefont {K.-M.}\ \bibnamefont {Hsieh}}, \bibinfo {author} {\bibfnamefont {B.-Y.}\ \bibnamefont {Wu}}, \bibinfo {author} {\bibfnamefont {Z.}~\bibnamefont {Niu}}, \bibinfo {author} {\bibfnamefont {F.}~\bibnamefont {Aziz}}, \bibinfo {author} {\bibfnamefont {Y.-H.}\ \bibnamefont {Huang}}, \bibinfo {author} {\bibfnamefont {P.}~\bibnamefont {Wen}}, \bibinfo {author} {\bibfnamefont {K.-T.}\ \bibnamefont {Lin}}, \bibinfo {author} {\bibfnamefont {Y.-H.}\ \bibnamefont {Lin}}, \bibinfo {author} {\bibfnamefont {J.}~\bibnamefont {Chen}}, \bibinfo {author} {\bibfnamefont {A.}~\bibnamefont {Kockum}}, \bibinfo {author} {\bibfnamefont {G.-D.}\ \bibnamefont {Lin}}, \bibinfo {author} {\bibfnamefont {Z.-R.}\ \bibnamefont {Lin}}, \bibinfo {author} {\bibfnamefont {Y.}~\bibnamefont {Lu}},\ and\ \bibinfo {author} {\bibfnamefont {I.-C.}\ \bibnamefont {Hoi}},\ }\bibfield  {title} {\bibinfo {title} {Group {Delay}
  {Controlled} by the {Decoherence} of a {Single} {Artificial} {Atom}},\ }\href {https://doi.org/10.1103/fkzb-fxv4} {\bibfield  {journal} {\bibinfo  {journal} {Physical Review Letters}\ }\textbf {\bibinfo {volume} {135}},\ \bibinfo {pages} {073601} (\bibinfo {year} {2025})}\BibitemShut {NoStop}%
\bibitem [{\citenamefont {Sinha}\ \emph {et~al.}(2020)\citenamefont {Sinha}, \citenamefont {Meystre}, \citenamefont {Goldschmidt}, \citenamefont {Fatemi}, \citenamefont {Rolston},\ and\ \citenamefont {Solano}}]{sinha2020non}%
  \BibitemOpen
  \bibfield  {author} {\bibinfo {author} {\bibfnamefont {K.}~\bibnamefont {Sinha}}, \bibinfo {author} {\bibfnamefont {P.}~\bibnamefont {Meystre}}, \bibinfo {author} {\bibfnamefont {E.~A.}\ \bibnamefont {Goldschmidt}}, \bibinfo {author} {\bibfnamefont {F.~K.}\ \bibnamefont {Fatemi}}, \bibinfo {author} {\bibfnamefont {S.~L.}\ \bibnamefont {Rolston}},\ and\ \bibinfo {author} {\bibfnamefont {P.}~\bibnamefont {Solano}},\ }\bibfield  {title} {\bibinfo {title} {Non-{M}arkovian collective emission from macroscopically separated emitters},\ }\href {https://doi.org/10.1103/PhysRevLett.124.043603} {\bibfield  {journal} {\bibinfo  {journal} {Physical review letters}\ }\textbf {\bibinfo {volume} {124}},\ \bibinfo {pages} {043603} (\bibinfo {year} {2020})}\BibitemShut {NoStop}%
\bibitem [{\citenamefont {Arranz~Regidor}\ and\ \citenamefont {Hughes}(2021)}]{arranz2021cavitylike}%
  \BibitemOpen
  \bibfield  {author} {\bibinfo {author} {\bibfnamefont {S.}~\bibnamefont {Arranz~Regidor}}\ and\ \bibinfo {author} {\bibfnamefont {S.}~\bibnamefont {Hughes}},\ }\bibfield  {title} {\bibinfo {title} {Cavitylike strong coupling in macroscopic waveguide {QED} using three coupled qubits in the deep non-{M}arkovian regime},\ }\href {https://doi.org/10.1103/PhysRevA.104.L031701} {\bibfield  {journal} {\bibinfo  {journal} {Physical Review A}\ }\textbf {\bibinfo {volume} {104}},\ \bibinfo {pages} {L031701} (\bibinfo {year} {2021})}\BibitemShut {NoStop}%
\bibitem [{\citenamefont {Breuer}\ \emph {et~al.}(2016)\citenamefont {Breuer}, \citenamefont {Laine}, \citenamefont {Piilo},\ and\ \citenamefont {Vacchini}}]{breuer2016colloquium}%
  \BibitemOpen
  \bibfield  {author} {\bibinfo {author} {\bibfnamefont {H.-P.}\ \bibnamefont {Breuer}}, \bibinfo {author} {\bibfnamefont {E.-M.}\ \bibnamefont {Laine}}, \bibinfo {author} {\bibfnamefont {J.}~\bibnamefont {Piilo}},\ and\ \bibinfo {author} {\bibfnamefont {B.}~\bibnamefont {Vacchini}},\ }\bibfield  {title} {\bibinfo {title} {Colloquium: Non-{M}arkovian dynamics in open quantum systems},\ }\href {https://doi.org/10.1103/RevModPhys.88.021002} {\bibfield  {journal} {\bibinfo  {journal} {Reviews of Modern Physics}\ }\textbf {\bibinfo {volume} {88}},\ \bibinfo {pages} {021002} (\bibinfo {year} {2016})}\BibitemShut {NoStop}%
\end{thebibliography}%

\clearpage
\newpage

\end{document}